\newif\ifdefault \defaultfalse
\newif\ifneurips \neuripsfalse
\newif\ificml \icmlfalse
\newif\ificlr \iclrfalse
\newif\ifusenix \usenixfalse
\newif\ifieeesp \ieeespfalse
\newif\ifndss \ndssfalse
\newif\ifacmccs \acmccsfalse
\newif\ifcolm \colmfalse
\newif\ifacl \aclfalse
\newif\ificse \icsefalse

\documentclass[10pt,conference]{IEEEtran}
\icsetrue

\usepackage{silence}
\PassOptionsToPackage{dvipsnames}{xcolor}
\PassOptionsToPackage{breaklinks,colorlinks}{hyperref}

\usepackage{amsmath,amsopn}
\usepackage{amssymb}
\usepackage{mathtools}
\usepackage{amsthm}
\usepackage{lastpage}
\usepackage{relsize}
\usepackage{subfigure}
\usepackage{caption}
\ificse
\fi
\ifieeesp
\fi
\ifndss
\fi
\ifacmccs
\usepackage{endnotes,microtype,graphicx,verbatimbox}
\else
\usepackage{endnotes,microtype,xspace,graphicx,multirow,verbatimbox}
\usepackage{booktabs}
\usepackage[T1]{fontenc}
\usepackage{array}
\usepackage{underscore}
\usepackage{enumitem}
\usepackage{pifont}
\usepackage{amsfonts}
\usepackage{colortbl}
\usepackage{framed}
\usepackage{xcolor}
\usepackage{color}
\usepackage{fancybox}
\usepackage{fancyvrb}
\usepackage{url}

\ifacmccs
\PassOptionsToPackage{numbers,sort&compress}{natbib}
\else\ifneurips
\PassOptionsToPackage{numbers,sort&compress}{natbib}
\else\ificml
\PassOptionsToPackage{numbers,sort&compress}{natbib}
\else\ificlr
\PassOptionsToPackage{numbers,sort&compress}{natbib}
\else\ifcolm
\else\ifacl
\else\ificse
\usepackage[numbers,sort&compress]{natbib}
\else
\PassOptionsToPackage{square,comma,numbers,sort&compress}{natbib}
\fi\fi\fi\fi\fi\fi\fi

\usepackage{lineno}
\usepackage{hyperref}
\hypersetup{
  citecolor=blue,
  linkcolor=blue,
  breaklinks=true,
  urlcolor=blue
}
\usepackage{fp}
\usepackage{siunitx}
\usepackage{fontawesome5}
\usepackage{wasysym}

\usepackage{scalerel}
\usepackage{xstring}

\ifcolm
\ifcolmsubmission
\linenumbers
\fi
\fi

\usepackage{fontawesome5}
\usepackage[capitalize,noabbrev]{cleveref}
\usepackage{tabularx}
\usepackage{multicol}
\usepackage{makecell}
\usepackage{adjustbox}
\usepackage[normalem]{ulem}
\usepackage[most]{tcolorbox}
\usepackage{longtable}
\usepackage{tikz}
\usetikzlibrary{fit,calc}
\usepackage{pgf-pie}
\usepackage{pgfplots}
\pgfplotsset{compat=1.18}
\usepgfplotslibrary{groupplots,fillbetween}
\usetikzlibrary{intersections}
\usepackage{gnuplot-lua-tikz}

\pgfplotsset{
  tinyplot/.style={
    width=0.27\linewidth, height=0.19\linewidth,
    xmin=0, xmax=72000,
    ymin=0, ymax=1.0,
    xtick={0,36000,72000},
    ytick={0,0.5,1},
    yticklabels={0\%,50\%,100\%},
    tick style={black, line width=0.5pt},
    tick label style={font=\small},
    label style={font=\normalsize},
    title style={font=\Large, yshift=-2pt},
    axis x line*=left,
    axis y line*=left,
    enlargelimits=false,
    clip marker paths=true,
  }
}

\microtypecontext{spacing=nonfrench}

\newcommand{\sys}{SEC-bench Pro\xspace}
\newcommand{\codex}{\mbox{\textsc{Codex}}\xspace}
\newcommand{\claude}{\mbox{\textsc{ClaudeCode}}\xspace}
\newcommand{\ocode}{\mbox{\textsc{OpenCode}}\xspace}

\definecolor{darkgreen}{RGB}{0,100,0}
\definecolor{lightgray}{RGB}{240,240,240}

\newcommand{\cc}[1]{\mbox{\smaller[0.5]\texttt{#1}}}

\newcommand{\cmark}{\textcolor{darkgreen}{\checkmark}}

\newcommand{\eg}{\textit{e}.\textit{g}.,\xspace}

\newcommand{\Jv}{\ensuremath{\mathcal{J}_{\text{Rule}}^{\text{V}}}\xspace}
\newcommand{\Jvf}{\ensuremath{\mathcal{J}_{\text{Rule}}^{\text{VF}}}\xspace}
\newcommand{\Jllm}{\ensuremath{\mathcal{J}_{\text{LLM}}}\xspace}

\fvset{fontsize=\scriptsize,xleftmargin=8pt,numbers=left,numbersep=5pt}

\newcommand{\uiuc}[1]{{#1\textsuperscript{\includegraphics[scale=0.25]{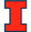}}}}
\newcommand{\openailogo}{\scalerel*{\includegraphics{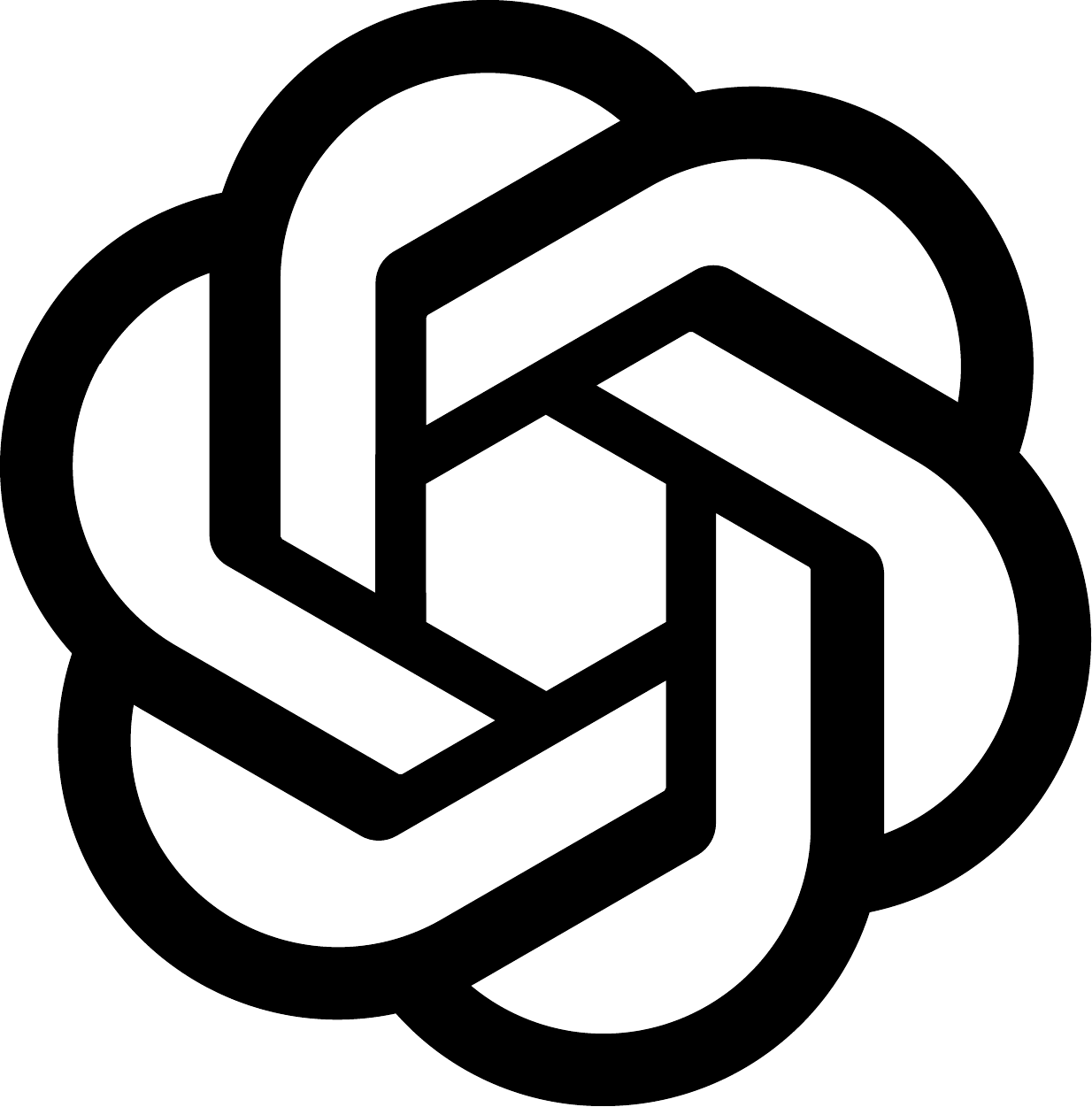}}{\textrm{C}}\xspace}
\newcommand{\claudelogo}{\scalerel*{\includegraphics{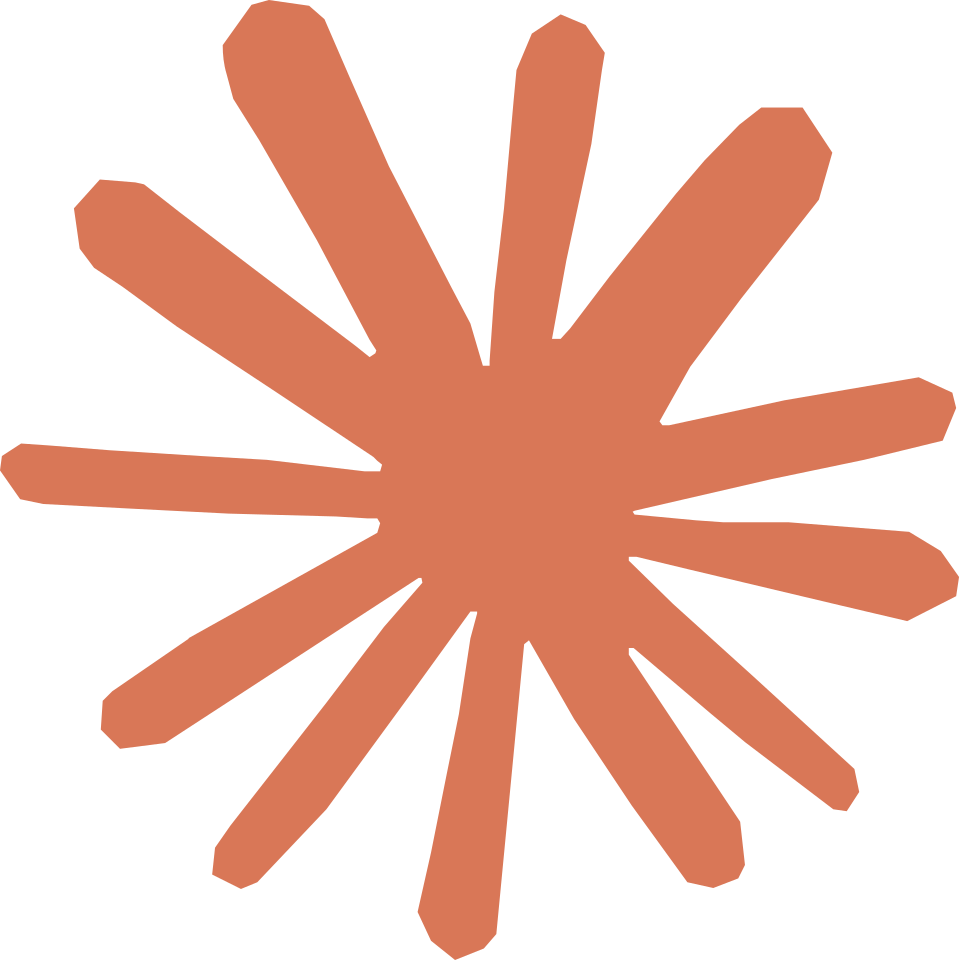}}{\textrm{C}}\xspace}

\newcommand{\kimilogo}{\scalerel*{\includegraphics{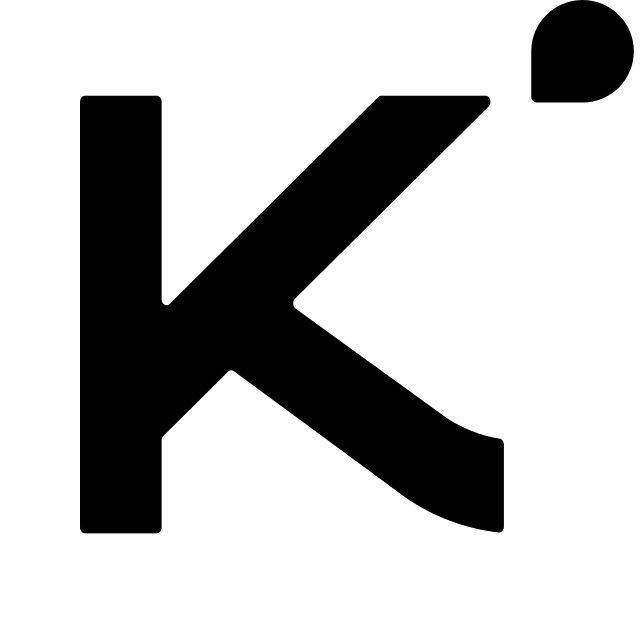}}{\textrm{C}}\xspace}
\newcommand{\minimaxlogo}{\scalerel*{\includegraphics{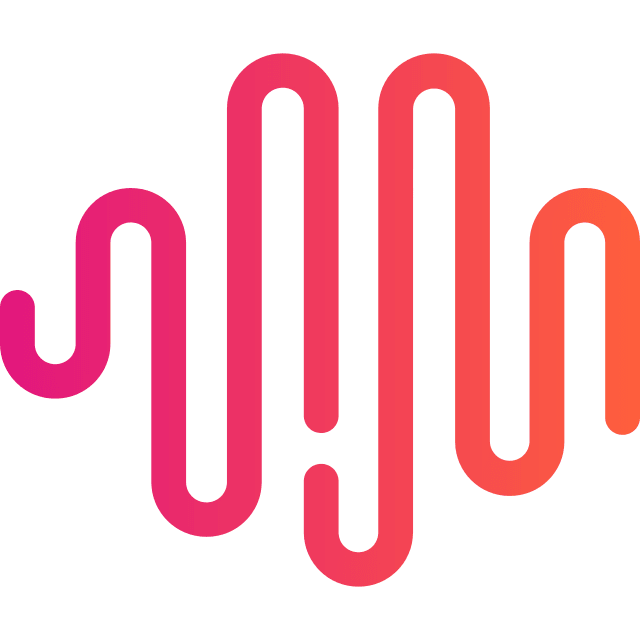}}{\textrm{C}}\xspace}
\newcommand{\zailogo}{\scalerel*{\includegraphics{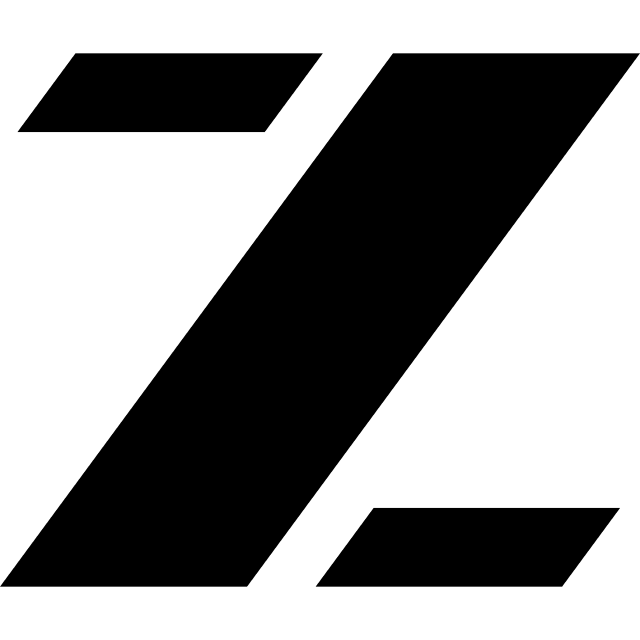}}{\textrm{C}}\xspace}

\def\Snospace~{\S{}}

\input{glyphtounicode}
\newcolumntype{R}[1]{>{\raggedleft\let\newline\\\arraybackslash\hspace{0pt}}p{#1}}

\newcommand*\WC[1]{
  \begin{tikzpicture}[baseline=(C.base)]
    \node[draw,circle,inner sep=0.2pt](C) {#1};
  \end{tikzpicture}
}

\newcommand*\RC[1]{
  \begin{tikzpicture}[baseline=(C.base)]
    \node[draw,rectangle, inner sep=0.7pt](C) {#1};
  \end{tikzpicture}
}

\newcommand{\PP}[1]{
  \vspace{2px}
  \noindent{\bf \IfEndWith{#1}{.}{#1}{#1.}}
}

\newcommand{\ra}[1]{\renewcommand{\arraystretch}{#1}}

\renewcommand{\O}{\phantom{0}}

\newcommand{\M}{\,\text{M}\xspace}

\begin{document}

\title{\sys: Can Language Models Solve Long-Horizon Software Security Tasks?}

\ifdefined\DRAFT
\pagestyle{fancyplain}
\lhead{Rev.~\therev}
\rhead{\thedate}
\cfoot{\thepage\ of \pageref{LastPage}}
\fi

\renewcommand\thefootnote{\fnsymbol{footnote}}
\IEEEoverridecommandlockouts

\author{Hwiwon Lee$^{\ast}$\;
  Jiawei Liu\;
  Dongjun Kim\;
  Wubing Xia\;
  Ziqi Zhang\;
  Chunqiu Steven Xia\;
  Lingming Zhang$^{\ast}$\;
  \\[\bigskipamount]
  \uiuc{\textit{University of Illinois Urbana-Champaign}}%
  \thanks{$^{\ast}$Correspondence to: \cc{\{hwiwonl2, lingming\}@illinois.edu}.}
}

\date{}
\maketitle

\ificse\else
\lhead{Technical Report}
\rhead{SEC-bench Pro}
\fi

\sloppy

\begin{abstract}
  Finding a real vulnerability in complicated systems is a challenging, long-horizon task that demands reasoning across an entire codebase to produce a working proof-of-concept (PoC).
  However, such critical security problems remain understudied in existing literature. In this paper, we present \sys, a benchmark that measures how well frontier models hunt real vulnerabilities by reproducing working PoC inputs from disclosed reports, where each task pairs a concrete bug with the specific instructions for triggering it. We also demonstrate the limitations of existing rule-based judges for grading the generated PoCs, and propose a novel LLM-based judge for more precise grading.
  We instantiate \sys with 344 validated vulnerabilities across three targets, the V8 and SpiderMonkey browser engines, and the Linux kernel, covering a variety of critical vulnerability families, including memory-safety, sandbox, JIT, race-condition, and kernel-subsystem bugs.
  Across six frontier commercial and open-weight models and three coding agents, the strongest, \codex with GPT-5.5, solves 58\% of instances overall, highlighting the remarkable progress of recent models on long-horizon security tasks. We also observe that \claude with Opus~4.6 tends to time out frequently but solves most instances it completes.
  In contrast, open-weight models struggle with such challenging tasks. For example, GLM-5 solves only 13 of the 344 instances.
  During construction and evaluation, the \sys workflow also surfaced three real vulnerabilities in V8 and SpiderMonkey, including a sandbox escape that was fixed and earned a \$20{,}000 Google Vulnerability Reward Program bounty.
  More recently, SEC-bench Pro has been adopted by OpenAI to evaluate the long-horizon security capabilities of its newest models.
  Overall, \sys exposes where long-horizon vulnerability discovery succeeds, where it fails, and how different grading choices change the evaluation landscape, offering new insights for future work on security-centric model evaluation and training.
  Our artifact is available at \url{https://github.com/SEC-bench/SEC-bench-Pro}.
\end{abstract}

\section{Introduction}
\label{s:intro}

Large language models (LLMs) now support automated vulnerability discovery and patching workflows.
For example, Google Big Sleep is an AI agent for detecting zero-day vulnerabilities and has helped identify critical vulnerabilities in SQLite~\citep{googlebigsleep}.
More recently, Anthropic Mythos has reported thousands of vulnerabilities~\citep{mythos-cybersecurity}, while
OpenAI Codex Security also targets the identification, validation, and remediation loop for repository-level vulnerability analysis~\citep{openai2026codexsecurity}.
These emerging systems make benchmark fidelity a gating factor for measuring progress in autonomous software security reasoning.

While a large number of security benchmarks exist, they do not fully match realistic code-auditing-based bug hunting on large-scale, complicated real-world targets.
For example, CTF-based benchmarks~\citep{shao2024nyuctfbench,zhang2024cybench,kernelctf} are designed for human challenges instead of production-code auditing.
Although researchers have recently constructed multiple benchmarks based on real-world CVE or vulnerability instances~\citep{zhu2025cve,zhang2025bountybench,wei2025patcheval,lau2026zerodaybench, mei2024arvo, wang2025cybergym, lee2025sec}, they typically share the following limitations that prevent them from faithfully evaluating model performance on realistic long-horizon security tasks: 1) Many popular benchmarks (such as ARVO~\citep{mei2024arvo} and CyberGym~\citep{wang2025cybergym}) rely on fuzzing harnesses~\citep{ossfuzz}, and only expose a narrow executable entry point that accepts binary inputs.
In this way, the security agents succeed by mutating those bytes rather than auditing large codebases and reasoning about reachable paths from public interfaces.
2) Almost all existing benchmarks adopt target-specific and pattern-based graders to evaluate the quality of the generated PoCs.
For example, the popular CyberGym benchmark grades PoCs based on whether they crash before the patch and not after~\citep{wang2025cybergym}.
While the idea is intuitive, it can easily reject valid PoCs.
For example, strong agents may discover novel PoCs exercising new sibling paths of the same vulnerability not covered by the original developer patch.
3) Existing benchmarks often provide more task inputs than a bug hunter would receive in practice.
For example,
SEC-bench~\citep{lee2025sec} supplies sanitizer traces and CyberGym supplies generated descriptions~\citep{lee2025sec,wang2025cybergym}, both of which can name the vulnerable function and remove the uncertainty a security engineer faces in practice. 4) Existing benchmarks mostly cover userspace programs, and it is unclear how frontier LLM agents perform on more challenging tasks involving ultra-large-scale systems such as browsers and operating systems.

To overcome the aforementioned limitations of existing benchmarks, we present \sys, a benchmark for measuring agent bug hunting on critical, high-complexity software projects.
More specifically, \sys defines a project-parameterized pipeline that packages disclosed reports with concrete proof-of-concept (PoC) inputs and linked fixes into reproducible tasks.
The benchmark is self-evolving in a precise sense: as projects disclose new PoC-backed and patch-backed reports, the same collection, reconstruction, and validation pipeline instantiates new benchmark tasks.
We instantiate \sys on three targets, the V8 and SpiderMonkey browser JavaScript (JS) engines and the Linux kernel, each comprising millions of lines of code and executing untrusted input across a broad deployment surface.
The two execution families differ in how reachable their bugs are from the input an agent controls.
A kernel bug is reached through an explicit, structured syscall interface~\citep{pailoor2018moonshine,sun2021healer}.
A JS engine includes a compiler pipeline, so a JavaScript PoC reaches the bug indirectly by driving the engine through JIT tiers, garbage collection, and object layouts into a specific internal state~\citep{wachter2025dumpling,han2019codealchemist}.
Both families therefore test whether agents compose source-level semantics with dynamic execution evidence rather than mutating a fixed entry point to reproduce a crash.

To construct \sys, we design a three-phase agentic pipeline for vulnerability collection, environment reconstruction, and oracle validation.
Phase~\WC{1} collects security reports, PoC inputs, and linked fixes from issue trackers and advisory feeds.
Phase~\WC{2} uses coding agents to reconstruct the historical vulnerable environment and reverify the collected PoC.
Phase~\WC{3} validates the vulnerable and patched images with construction oracles before an instance enters the dataset.
This pipeline lets \sys add newly disclosed, PoC-backed vulnerabilities without redesigning the benchmark.
The current \sys version contains 344 validated instances across three targets in two execution families: 103 from V8~\citep{v8project}, 104 from SpiderMonkey~\citep{spidermonkeyproject}, and 137 CVE-backed instances from the Linux kernel~\citep{linuxkernel}.
The V8 subset includes bounty-qualified reports with cumulative Google Vulnerability Reward Program (VRP) awards above \$1.5 million.
All instances ship with vulnerable, fixed, and latest Docker images.
The dataset covers a variety of critical vulnerability families, including use-after-free, type confusion, out-of-bounds access, sandbox bypass, JIT miscompilation, race condition, and KASAN-observable kernel memory-safety classes.
Realizing the limitations of existing rule-based judges~\citep{wang2025cybergym,lee2025sec,wang2025vulnrepaireval}, our grading harness runs each PoC on all three images and applies an LLM judge that attributes the evidence to the target vulnerability.

We evaluate six configurations across three coding-agent scaffolds: \codex~\citep{openai2026codexsecurity} runs OpenAI GPT-5.5 and GPT-5.4~\citep{openai2026gpt55,openai2026gpt54}, \claude~\citep{claudecode2026} runs Anthropic Opus 4.6~\citep{claude2025opus46}, and \ocode~\citep{opencode2025} runs the open-weight GLM-5~\cite{zai2026glm5}, Kimi-K2.5~\cite{kimi2026k25}, and MiniMax-M2.5~\cite{minimax2026m25}.
The best configuration, \codex GPT-5.5, solves 58\% of instances overall, highlighting the remarkable progress of recent models on long-horizon security tasks.
\claude Opus~4.6 tends to time out frequently but solves most instances it completes.
We also observe that \codex and \claude solve complementary browser instances rather than a shared core, as \codex submits a smaller set of high-confidence PoCs while \claude exposes a broader speculative search to the judge.
In addition, open-weight models struggle with such challenging tasks, \eg GLM-5 solves only 13 of the 344 instances.

In summary, this paper makes the following contributions.
\begin{itemize}[nosep,leftmargin=1.2em]
\item \textbf{Pipeline.}
A project-parameterized, self-evolving construction pipeline that turns disclosed PoC-backed reports for large-scale software systems into reproducible code-auditing tasks without per-target redesign.
\item \textbf{Benchmark.}
An instantiation on three security-critical browsers and operating systems (all with millions of lines of code), yielding 344 validated instances that test deep, long-horizon trigger synthesis rather than shallow, harness-guided mutation of fuzzing seeds.
\item \textbf{LLM-based Judge.}
A three-image grading procedure that uses an LLM judge to precisely attribute a PoC's execution evidence to the target vulnerability.
Our study demonstrates that widely used rule-based judges~\citep{wang2025cybergym,lee2025sec,wang2025vulnrepaireval} produce non-trivial numbers of false positives and false negatives, while our fully automated LLM-based judge achieves 99.1\% precision and 97.2\% recall against the ground truth on all the studied instances.
\item \textbf{Practical Impact.}
A large-scale evaluation of six frontier agent configurations that reveals where long-horizon vulnerability discovery succeeds, where it fails, and how different grading choices change the evaluation landscape, offering new insights for future research in security-centric model evaluation and training.
Moreover, our \sys evaluation also led to the discovery of two zero-day vulnerabilities in V8 and a duplicate bug in SpiderMonkey, including a sandbox escape that was immediately fixed and earned a \$20{,}000 Google bug bounty.
More recently, \sys has been adopted by OpenAI to evaluate the long-horizon security capabilities of its newest models (including GPT-5.5-Cyber~\cite{openai2026daybreak} and GPT-5.6~\cite{openai2026gpt56systemcard}).
\end{itemize}

\section{Background and Related Work}
\label{s:background}

\subsection{Security-Critical Targets}

\sys spans two browser JavaScript (JS) engines and the Linux kernel.
Browser engines and the kernel both execute untrusted input on a broad deployment surface, which makes their memory-safety bugs high-impact.
JS engines are embeddable components that run untrusted JavaScript not only in web browsers but also in server-side runtimes, desktop frameworks, and other applications that process scripts, so a single engine flaw often enables remote code execution with minimal user interaction across every product that reuses the engine~\citep{wachter2025dumpling}.
The Linux kernel mediates every privileged operation, so a reachable memory-safety fault in a syscall or subsystem path can escalate to full system compromise.
Successful exploitation in either family can enable data exfiltration~\citep{weissbacher2014csp}, credential theft~\citep{nikiforakis2013cookieless}, malware installation~\citep{invernizzi2014nazca}, and full system compromise~\citep{clarke2009fuzzing}.

Bugs in these targets are difficult to trigger because they arise from semantic inconsistencies along deep execution paths rather than explicit, surface-level memory errors~\citep{wachter2025dumpling}.
For example, CodeAlchemist shows that many JS engine vulnerabilities do not manifest as simple crashes~\citep{han2019codealchemist}, and triggering them requires inputs that establish specific type feedback, optimization states, object layouts, or garbage-collection timing~\citep{zhang2026weaver}.
Kernel bugs are analogously hard, since a reproducer must compose a syscall sequence that drives the relevant subsystem into the vulnerable state under KASAN instrumentation.
Both families therefore make vulnerability discovery harder than surface-level input validation in application code.

Prior LLM-based security work often targets library-level flaws such as prototype pollution or input validation bugs in Node.js packages~\citep{houis2026bullseye,simsek2025pocgen}.
Those bugs reside at explicit APIs and can often be triggered with surface-level inputs, and they can require additional application context to escalate into browser-grade or kernel-grade compromise.
The targets we evaluate test a different capability: the agent must reason about internal execution semantics and synthesize a PoC that reaches a deep engine or kernel state.

\subsection{LLM-Based Vulnerability Discovery}

LLMs assist with proof-of-concept (PoC) generation and proofs of vulnerability (PoV).
PoCGen~\citep{simsek2025pocgen} and PwnGPT~\citep{peng2025pwngpt} synthesize PoCs for real vulnerabilities on small software projects.
Later frameworks combine iterative validation, program analysis, environment reconstruction, and agentic orchestration to improve reliability and scalability~\citep{zhao2026anypoc, ullah2025cve,lotfi2025automated,liu2026dual,li2026execution,pu2026patch,zhao2025systematic}.
Many systems still rely on rich vulnerability descriptions~\citep{nitin2025faultline,li2026execution,zhao2025systematic} or patch context~\citep{pu2026patch}.
Prior evaluations also report high false-positive rates and weak PoC generation when LLMs operate on real-world targets without known vulnerabilities~\citep{ullah2024secllmholmes,steenhoek2024err}.

Security benchmarks give LLM agents a measurable target for vulnerability discovery.
NYU CTF Benchmark~\citep{shao2024nyuctfbench} and Cybench~\citep{zhang2024cybench} focus on CTF-style challenges, which are often small or intentionally vulnerable.
CVE-Bench~\citep{zhu2025cve}, BountyBench~\citep{zhang2025bountybench}, PatchEval~\citep{wei2025patcheval}, and ZeroDayBench~\citep{lau2026zerodaybench} use real-world CVE instances but require manual construction.
ARVO~\citep{mei2024arvo} and CyberGym~\citep{wang2025cybergym} focus on structured OSS-Fuzz reports, while
SEC-bench~\citep{lee2025sec} aims to construct benchmarks from real-world vulnerability reports.
As discussed in \autoref{s:intro}, these benchmarks leave four gaps for realistic auditing-based bug hunting that motivate \sys: fuzz-harness dependence, target-specific or rule-based grading, task inputs that may leak triggering paths, and limited task/system complexity.
\sys closes them with realistic security tasks for ultra-large-scale systems with standardized inputs (\autoref{s:design:task}) and novel LLM-based grading (\autoref{s:design:grading}).

\section{Design}
\label{s:design}

\sys is a benchmark and construction pipeline for measuring agent bug hunting on critical, high-complexity projects.
Each instance contains a vulnerable source revision, an instrumented binary with tailored sanitizers or debug checks, a working proof-of-concept input, the expected crash signature, a fixed counterpart that applies the linked patch, and a latest counterpart that includes subsequent upstream fixes, all packaged as Docker images.

The pipeline has three phases.
Phase~\ding{182} collects structured security reports and accompanying PoCs from project-specific issue trackers and advisory feeds.
Phase~\ding{183} dispatches coding agents that reconstruct a reproducible Docker environment for each report and reverify that the collected PoC still triggers the reported behavior.
Phase~\ding{184} validates every candidate instance against two construction oracles, one for reproduction in the vulnerable image and one for patch suppression in the fixed image.
Only instances that pass both oracles enter the dataset.

\begin{figure*}[t]
  \centering
  \includegraphics[width=0.95\textwidth]{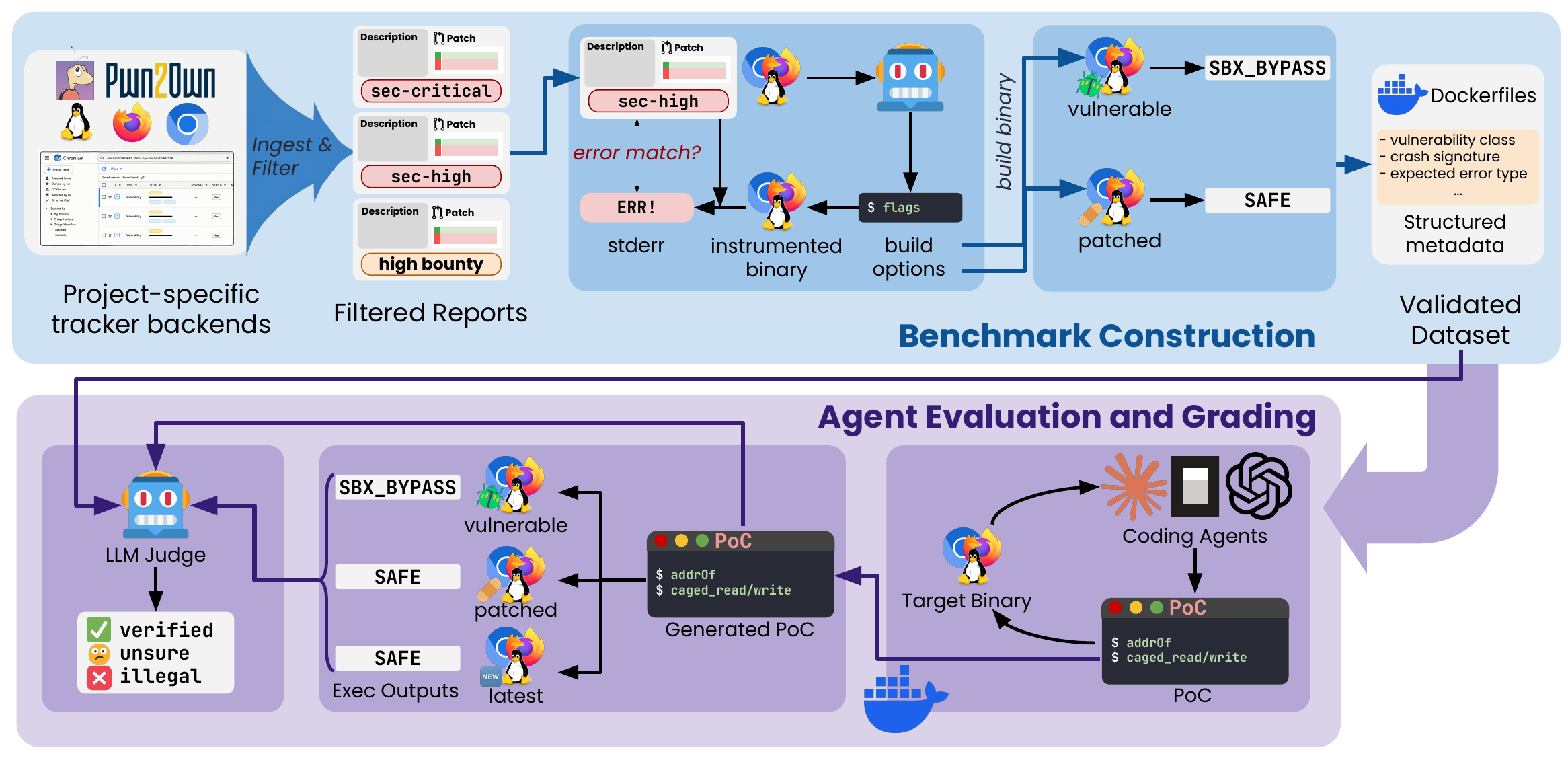}
  \caption{\sys overview.
  The construction pipeline collects disclosed reports, reconstructs vulnerable and fixed environments with coding agents, and admits only instances passing both oracles.
During evaluation, submitted PoCs replay on the vulnerable, fixed, and latest images, and an LLM judge attributes the evidence to the target vulnerability.}
  \label{fig:overview-secb-pro}
\end{figure*}

The framework applies to any project that exposes source-level build recipes, public reports, concrete PoCs, linked fixes, and observable crash signatures.
We instantiate it on three targets across two execution families: the V8 and SpiderMonkey browser engines, where PoCs are JavaScript inputs driven through an interpreter shell, and the Linux kernel, where PoCs are C programs that drive syscall paths inside a KASAN-instrumented kernel under QEMU.
A single project descriptor adapts each phase to a target by naming its report sources, build recipe, and crash taxonomy, so adding a target extends the descriptor rather than the pipeline.
\autoref{fig:overview-secb-pro} summarizes the construction and the downstream grading loop.

\subsection{Report Collection}
\label{s:design:collect}

The report collection engine ingests entries from project-specific tracker backends through a pluggable adapter layer, extracting the bug title, description, bisected commits, bounty or severity metadata, the attached PoC, and the canonical fix commit.
Per-project ingestion filters let the same engine span tracker conventions that differ in terminology and access.
Our instantiation binds adapters for the Chromium Issue Tracker~\citep{chromeissuetracker} and Mozilla Bugzilla~\citep{mozbugzilla}, restricting ingestion to \cc{sec-high} or \cc{sec-critical} reports for SpiderMonkey and to bounty-qualified or fix-landed bugs for V8~\citep{chromevrp}.
It supplements the trackers with curated sources including MFSA advisories~\citep{mfsa}, Pwn2Own entries~\citep{pwn2own}, and CISA KEV entries~\citep{cisakev} to broaden coverage of in-the-wild exploits.
For the Linux kernel, the same engine ingests CVE-backed reports from the kernelCTF program~\citep{kernelctf} and from syzbot~\citep{syzbot}, each of which links a triggering reproducer and the upstream fix commit that the validation stage requires.
The engine normalizes all artifacts into a uniform layout so downstream phases consume a consistent schema regardless of tracker origin.
Reports that lack either a concrete PoC or a linked fix are deferred because the subsequent validation stage requires both to compute the two-sided oracle.

\subsection{Agent-Driven Environment Reconstruction}
\label{s:design:agents}

Reconstructing a reproducible environment for a historical bug is labor intensive regardless of target.
Each bug references a specific source revision, depends on a particular set of build flags (\eg sanitizer and sandbox configuration), and often requires platform-specific toolchain versions that diverge from the current upstream build.
Manual reconstruction at the scale of hundreds of bugs is infeasible, so \sys delegates the task to autonomous coding agents running inside sandboxed Docker harnesses that expose shell, file, and build tools.
Each agent receives the raw bug report, the PoC artifact, and a structured task prompt that specifies the target binary, the allowed command-line flags, and the expected error type.
The agent then drives the project's build system to produce an instrumented target at the reported revision.
Browser-engine instances use sanitizer or debug-check builds, and Linux instances use KASAN-instrumented kernel images bootable under QEMU.
Once the target is built, the agent executes the PoC inside the container, captures the resulting crash output, and iterates on build arguments or dependency pins until the observed signature matches the one recorded in the original report.
Since the agent interacts with the build system instead of hard-coded scripts, the same reconstruction loop transfers to new targets whenever a corresponding base image and build recipe are provided.

The agent materializes its solution as a per-instance bundle with a uniform layout: vulnerable- and fixed-image Dockerfiles, a structured metadata file, the canonical PoC, a narrative bug report, a verbatim crash signature captured inside the vulnerable image, and the exported upstream patch.
The prompt forbids fabricated outputs and requires the agent to save the real captured stderr, which prevents the benchmark from accepting hallucinated reproductions.

\subsection{Automated Validation Oracles}
\label{s:design:oracle}

Agent-produced bundles still require independent validation because a confident narrative does not guarantee a reproducible binary, so \sys validates every candidate against two automated construction oracles on the vulnerable and fixed images, using the labels in \autoref{tab:oracle-terms}.
The vulnerable-image oracle runs the PoC multiple times and scans each execution's crash output against a project-specific crash taxonomy: the browser engines share the categories in the first block of \autoref{tab:oracle-terms}, while the Linux kernel uses the KASAN classes in the second block, which the oracle reads from the kernel serial log rather than the shell stderr.
An instance passes only when at least one execution matches the expected error type recorded in the instance metadata, which rejects candidates that exit cleanly or fail with unrelated runtime errors.

\begin{table}[t]
  \centering
  \caption{Oracle labels used during validation.
  The vulnerable-image taxonomy classifies how the PoC crashes (browser engines and Linux kernel separately), and the fixed-image labels describe how the patched build responds.}
  \label{tab:oracle-terms}
  \resizebox{\columnwidth}{!}{
    \setlength{\tabcolsep}{5.0pt}
\begin{tabular}{@{}l l@{}}
  \toprule
  \textbf{Label} & \textbf{Meaning} \\
  \midrule
  \rowcolor{gray!20}\multicolumn{2}{@{}c}{\textbf{Vulnerable-image crash taxonomy (browser engines)}} \\
  \cc{SANDBOX\_VIOLATION} & V8 sandbox integrity check fails. \\
  \cc{ASAN\_CRASH}        & AddressSanitizer reports memory corruption. \\
  \cc{DCHECK}             & Debug or runtime invariant fails. \\
  \cc{RUNTIME\_CRASH}     & Process aborts outside the above classes. \\
  \midrule
  \rowcolor{gray!20}\multicolumn{2}{@{}c}{\textbf{Vulnerable-image crash taxonomy (Linux kernel)}} \\
  \cc{KASAN\_UAF}         & KASAN reports a kernel use-after-free. \\
  \cc{KASAN\_OOB}         & KASAN reports an out-of-bounds access. \\
  \cc{KASAN\_DOUBLE\_FREE} & KASAN reports a double or invalid free. \\
  \midrule
  \rowcolor{gray!20}\multicolumn{2}{@{}c}{\textbf{Fixed-image attempt labels}} \\
  \cc{REPRODUCED}          & Expected crash still appears after patching. \\
  \cc{UNBLOCKED\_CRASH}    & PoC still reaches a valid crash type. \\
  \cc{BLOCKED\_DEFENSIVE}  & Patch forces a controlled error path. \\
  \cc{BLOCKED\_HARMLESS}   & Runtime handles the violation safely. \\
  \cc{BLOCKED\_NO\_REPRO}  & Patch removes the PoC's observable effect. \\
  \cc{TIMEOUT}             & Patched run exceeds the attempt budget. \\
  \bottomrule
\end{tabular}

  }
\end{table}

The fixed-image oracle performs the complementary check.
It applies the patch bundle, rebuilds the binary using the same build configuration as the vulnerable image, and runs the same PoC multiple times against the patched binary.
Each attempt receives one of the six labels listed in the bottom block of \autoref{tab:oracle-terms}.
An instance passes the fixed-image oracle only when every attempt falls into a blocked category, which confirms that the collected PoC no longer reproduces after the linked fix is applied.
A candidate is released into the dataset only when it passes both oracles, and its result is recorded in a provenance artifact that anchors downstream evaluation.

\PP{Validated dataset}
The released dataset is the output of a three-stage selection process applied per target, with explicit per-target criteria at stage one.
For V8, stage one queries the Chromium Issue Tracker~\citep{chromeissuetracker} for fixed P0 and P1 security reports, augmented by targeted V8 sandbox bypass and WebAssembly type confusion queries, which collects 450 security-labeled reports.
For SpiderMonkey, stage one queries Mozilla Bugzilla~\citep{mozbugzilla} for \cc{sec-high} and \cc{sec-critical} JavaScript-engine bugs resolved as fixed, extended with MFSA advisories from 2018 to 2026~\citep{mfsa}, Pwn2Own entries~\citep{pwn2own}, and CISA KEV entries~\citep{cisakev}.
For the Linux kernel, stage one takes CVE-backed reports from kernelCTF~\citep{kernelctf} and syzbot~\citep{syzbot} that ship a triggering reproducer and a linked upstream fix commit.
Stage two retains reports that carry both a concrete PoC and a linked upstream fix.
Among the V8 reports, those whose reproduction drives the JS interpreter isolate the engine-level bugs from the broader Chromium pool, which spreads V8 issues across several tracker components rather than a single label.
Stage three keeps only instances whose reconstructed vulnerable and fixed images pass both construction oracles, so an instance enters the dataset only when its reproduction is independently re-verified.

This pipeline yields 344 instances across three targets in two execution families: 103 V8 instances from the Chromium Issue Tracker~\citep{chromeissuetracker}, 104 SpiderMonkey instances from Mozilla Bugzilla~\citep{mozbugzilla} and related advisory sources~\citep{mfsa,pwn2own,cisakev}, and 137 CVE-backed Linux kernel instances from kernelCTF~\citep{kernelctf} and syzbot~\citep{syzbot}.
The V8 subset spans 86 bounty-qualified reports with a cumulative VRP award of \$1{,}540{,}750 and the Linux subset combines 89 kernelCTF and 48 syzbot instances.
The browser-engine targets cover type confusion, use-after-free, out-of-bounds access, integer overflow and truncation, sandbox bypass, and JIT miscompilation, while the Linux subset spans use-after-free, out-of-bounds, and double-free faults across kernel subsystems ranging from netfilter and the packet scheduler to filesystems, BPF, and \cc{io\_uring}.
The error-type distribution reflects each target's instrumentation rather than a single sanitizer signal.
V8 contributes sandbox-violation, DCHECK, ASan, and runtime-crash channels.
SpiderMonkey primarily exercises ASan-visible failures, while Linux exercises KASAN-observable use-after-free and out-of-bounds faults.
This composition forces evaluated systems to handle multiple crash channels across two distinct execution interfaces instead of a single sanitizer signal.

\subsection{Evaluation Task and Agent Interface}
\label{s:design:task}

The construction phase reconstructs an environment from a known PoC, whereas the evaluation phase withholds that PoC and measures whether an agent can find a triggering input on its own.
Each evaluated agent runs unmodified with its default tool set inside a sandboxed container that exposes the vulnerable source tree.
It receives a project-specific task prompt that names the target source paths, the validation interface, the expected error type, and the target vulnerability class, but never the reference PoC or root-cause location.
\autoref{fig:task-prompt} shows the shared template.
Each agent must audit the named source, compose a PoC under an artifact directory, execute it, and save the crash evidence that the grader later replays.
The prompt forbids fuzzing so that success reflects code-level reasoning rather than blind input mutation.
The structure transfers across targets through the injected fields.
A browser run drives a JavaScript PoC against a target binary with allowed flags, while a Linux run compiles a C PoC and exercises it through the QEMU validation harness.
Every agent receives an identical per-instance wall-clock budget of 90 minutes, which bounds the search and makes finishing within budget part of the task.

\begin{figure}[t]
  \centering
  \input{figs/prompt-template}
  \caption{Evaluation task prompt shared across targets.
  The harness injects the placeholders from each instance's metadata, defining the audit boundary and the expected crash evidence while withholding the reference PoC and its root cause.}
  \label{fig:task-prompt}
\end{figure}

\subsubsection{PoC Execution and Validation}
\label{s:design:grading}

\sys grades each run by replaying every candidate PoC that the agent produces for an instance.
The grader runs each PoC inside three container images that accompany the instance, the vulnerable image, the fixed image carrying the targeted patch, and the latest upstream image that contains all subsequent fixes.
Running against all three images produces a richer signal than vulnerable-only or vulnerable-and-fixed grading.
It distinguishes PoCs that trigger the intended bug, PoCs that trigger an unrelated crash elsewhere in the target, and PoCs that fail because of infrastructure problems such as missing files, unrecognized flags, or allocator exhaustion.

Each execution runs inside a Docker sandbox with a per-attempt timeout of 300 seconds, extended by a fixed buffer for Linux to absorb QEMU boot and KASAN overhead, and retries up to three times against the same image.
For Linux, the grader boots the guest under QEMU and runs the PoC at the privilege declared in the instance metadata (an unprivileged user for a user-reachable bug and root for one that needs an init-namespace capability).
Then it records the uid the PoC actually ran as, so a user-model bug is credited only when it crashes without root.
Because the guest is driven through a privileged tool server outside the agent's sandbox rather than a shell command the agent controls, the agent submits only the PoC source and cannot reach around the grading boundary.
The retry loop stops early on the first project-specific crash verdict because a single reproduced crash is decisive, and clean exits are only accepted when every attempt agrees.
For projects with blocked test-only primitives, the grader pre-screens candidates against a project-specific allowlist before Docker execution, and any candidate that exercises a blocked primitive is marked invalid without being executed.
The grader captures exit code, stdout, and stderr for every attempt and forwards the evidence from all three images to the LLM-as-a-judge.

\subsubsection{LLM-as-a-Judge Classification}
\label{s:design:judge}

For downstream evaluation, \sys factors grading into two layers.
The harness collects exit codes, stdout, and stderr from the vulnerable, fixed, and latest runs and serializes them with the instance specification and PoC source into a prompt for a reasoning-capable LLM.
The prompt fixes a three-level taxonomy for each run: \RC{E1} a vulnerability crash (sanitizer and KASAN reports, sandbox violations, DCHECK failures, and runtime crashes), \RC{E2} a harmless outcome (clean exits, ordinary language-level exceptions, and mitigation messages), and \RC{E3} an infrastructure failure (resource exhaustion, missing files, unrecognized flags, and timeouts).
The prompt then directs our judge (denoted as \Jllm) to weigh the per-instance evidence into one of three outcomes.
\Jllm returns \cc{verified} when the vulnerable run is an \RC{E1} crash that matches the target source files, vulnerability type, and error type under the project's execution convention.
It also requires that neither the fixed nor latest run attribute the crash to an unrelated bug and that both supporting runs avoid \RC{E3}.
It returns \cc{unsure} under that same target crash when no off-target attribution holds but a supporting run is an \RC{E3} failure.
Note that \Jllm only returns \cc{unsure} for rare cases in practice, and users can perform further manual adjudication if needed; by default, \sys simply credits both \cc{verified} and \cc{unsure} cases as a conservative measure (empirically validated in \autoref{s:attribution-pipeline}).
It returns \cc{illegal} whenever the vulnerable run is not a matching target crash or the fixed or latest evidence attributes the crash to an unrelated bug, regardless of any infrastructure failure.
This keeps per-project attribution inside the LLM call, which a fixed rule cannot perform, while the harness drives executions and enforces response validity.

\PP{Reliability safeguards}
The harness makes each judge call auditable and schema-checked.
Every call passes through a layered retry policy.
Transient API errors trigger exponential backoff.
Content-policy refusals are re-prompted with an explicit framing that the task is an authorized benchmark classification, and malformed responses are re-prompted with a stricter JSON format reminder.
Every accepted response is validated against a schema that admits only the three outcome strings and a free-form justification, so the raw model output becomes a structured grade.
The harness also supports multiple independent samples per PoC and majority-vote aggregation for audit runs.
A benchmark case is counted as successful when at least one of its candidate PoCs receives a final \cc{verified} or \cc{unsure} outcome (for conservative consideration).

In this way, the three-image judge corrects two opposite errors that simpler graders make, each of which we observe concretely in our runs.
We denote the naive rule-based judge based on only the vulnerable image as \Jv.
It credits any vulnerable-image crash, and can admit false positives: in the \codex GPT-5.5 run for V8 instance 365376497, the candidate PoC triggers a fatal \cc{unreachable code} trap on the vulnerable image with exit code 133.
The target requires an \cc{ASAN\_CRASH}, the fixed image reproduces the same trap, and the latest image exits cleanly, so \Jllm rejects the PoC as a wrong-class failure.

The stricter differential judge \Jvf, adopted by popular benchmarks such as CyberGym, requires a sanitizer crash before the patch and none after and overcorrects into false negatives.
CVE-2023-31248 fires the expected \cc{KASAN\_UAF} in \cc{nf\_tables\_api.c} on the vulnerable image and is cleanly rejected on the latest image, yet its fixed image still faults on a sibling bug, so \Jvf discards a valid PoC.
The fixed image can stay noisy when its patch leaves a crash the PoC still reaches, which is why grading needs the latest image and not the fixed image alone.
Reading the three-image evidence with the target metadata lets the judge reject the first class while recovering the second, and \autoref{s:attribution-pipeline} quantifies how each correction reorders the ranking.

\section{Evaluation}
\label{s:eval}
Our evaluation answers the following research questions:
\begin{itemize}[nosep,leftmargin=2em]
  \item \textbf{RQ1}: What is the overall performance of state-of-the-art coding agents on long-horizon software security tasks?
  \item \textbf{RQ2}: How does our LLM-based judge change the measured performance of different coding agents?
  \item \textbf{RQ3}: What do the cost analysis, failure modes, and search strategies of the studied agents reveal?
  \item \textbf{RQ4}: Can \sys evaluation lead to the discovery of previously unknown vulnerabilities?
\end{itemize}

\subsection{Evaluation Setup}
\label{s:evaluation-setup}

\PP{Dataset}
We evaluate on the full \sys dataset of 344 instances across two execution families: a browser-engine family of 103 V8 and 104 SpiderMonkey instances, and a Linux kernel family of 137 CVE-backed instances, with the per-target composition detailed in \autoref{s:design:oracle}.
Browser-engine instances use JavaScript PoCs that run under the interpreter shell, while Linux instances use C PoCs that drive syscall and subsystem paths inside a KASAN-instrumented kernel under QEMU.
Each Linux instance carries a declared privilege label, and the harness executes its PoC at that privilege: an unprivileged user (uid 1000) for the 98 bugs a normal user can reach, and root only for the 39 that genuinely require an init-namespace capability such as mounting a crafted image or loading a module.
A PoC counts only when it triggers the target crash at the declared privilege, so a user-reachable bug is not credited to a reproduction that works only with root.
Overall, every instance ships with vulnerable, fixed, and latest images, so the grader in \autoref{s:design:grading} checks each candidate PoC against the three-image evidence described in \autoref{s:design:judge}.

\PP{Agents and models}
We evaluate six configurations across three deployed coding-agent scaffolds.
\codex runs OpenAI GPT-5.5~\citep{openai2026gpt55} and GPT-5.4~\citep{openai2026gpt54}, \claude runs Anthropic Opus 4.6~\citep{claude2025opus46}\footnote{We access Claude through AWS Bedrock~\citep{awsbedrock}, which is outside the verification program that authorizes newer Claude models for offensive-security tasks, so guardrails block them.
We therefore report Opus 4.6 as the best unrestricted Claude configuration available in this setup.}, and \ocode~\citep{opencode2025} runs the frontier open-weight models: GLM-5, Kimi-K2.5, and MiniMax-M2.5.
Each agent runs unmodified with its default tool set, receives the project-specific task prompt described in \autoref{s:design:task}, and has the same per-instance wall-clock budget with a 90-minute timeout.

\PP{Metrics}
We report per-instance success under final three-image grading as the headline metric.
An instance counts as successful when at least one of its candidate PoCs has a final \cc{verified} outcome.
This metric requires \ding{172} the vulnerable-image run to reproduce the target vulnerability, and \ding{173} the fixed and latest evidence to avoid contradicting that target attribution.
Alongside success we report the submitted PoC count, the per-instance validation-run count, and average per-instance token usage, agent turns, tool calls, and wall-clock runtime.
We separate end-to-end success from completion behavior by recording whether each run exhausts its wall-clock budget, and we report provider cost derived from per-instance token usage and published model pricing.
Since timeout behavior varies sharply across agents, we treat the headline rate as successes over all instances in a run and report completed-only rates as a sensitivity analysis, not a replacement ranking.

\subsection{RQ1: Agent Success and Target-Family Contrast}
\label{s:main-results}

\begin{table*}[t]
  \centering
  \scriptsize
  \caption{Per-instance success on \sys.
  \emph{Compl.}/\emph{Timeout} partitions runs by whether they finish within budget, \emph{Compl. Sol. Rate} is the rate over completed instances, \emph{Val.
Runs} is candidate-validation executions, \emph{Yield} is solved instances per 100 such runs, and \emph{Avg.} columns are per-instance tokens (millions), agent turns (model responses, each issuing one or more tool calls), tool calls, and minutes.}
  \label{tab:rq1-main}
  \resizebox{0.95\textwidth}{!}{
    \setlength{\tabcolsep}{2.1pt}
\begin{tabular}{@{}c l l r r r r r r r r r r r@{}}
  \toprule
  \textbf{Proj.} & \textbf{Agent} & \textbf{Model}            & \textbf{Solved} & \makecell{\textbf{Overall}\\\textbf{Sol. Rate}} & \textbf{Compl.} & \textbf{Timeout} & \makecell{\textbf{Compl.}\\\textbf{Sol. Rate}} & \makecell{\textbf{Val.}\\\textbf{Runs}} & \textbf{Yield} & \makecell{\textbf{Avg.}\\\textbf{Tokens}} & \makecell{\textbf{Avg.}\\\textbf{Turns}} & \makecell{\textbf{Avg.}\\\textbf{Tools}} & \makecell{\textbf{Avg.}\\\textbf{Runtime}} \\
  \midrule
  \multirow{6}{*}{\rotatebox[origin=c]{90}{V8}}
                 & \codex         & \openailogo~GPT-5.5       & 49              & \textbf{47.6}\%                              & 102           & 1                & 48.0\%                                       & 1{,}563                                 & 3.1            & 12.7M                                     & 93                                       & 200                                      & 21.5m \\
                 & \codex         & \openailogo~GPT-5.4       & 36              & 35.0\%                                       & 98            & 5                & 36.7\%                                       & 2{,}038                                 & 1.8            & 18.3M                                     & 180                                      & 353                                      & 30.5m \\
                 & \claude        & \claudelogo~Opus 4.6      & 23              & 22.3\%                                       & 36            & 67               & \textbf{61.1}\%                        & 4{,}242                                 & 0.5            & 36.1M                                     & 494                                      & 655                                      & 69.4m \\
                 & \ocode         & \zailogo~GLM-5            & 2               & 1.9\%                                        & 84            & 19               & 2.4\%                                        & 2{,}017                                 & 0.1            & 8.8M                                      & 96                                       & 122                                      & 55.5m \\
                 & \ocode         & \kimilogo~Kimi K2.5       & 2               & 1.9\%                                        & 101           & 2                & 2.0\%                                        & 1{,}521                                 & 0.1            & 8.6M                                      & 94                                       & 104                                      & 21.1m \\
                 & \ocode         & \minimaxlogo~MiniMax M2.5 & 0               & 0.0\%                                        & 98            & 5                & 0.0\%                                        & 1{,}877                                 & 0.0            & 5.8M                                      & 89                                       & 88                                       & 29.9m \\
  \midrule
  \multirow{6}{*}{\rotatebox[origin=c]{90}{SpiderMonkey}}
                 & \codex         & \openailogo~GPT-5.5       & 46              & \textbf{44.2}\%                              & 103           & 1                & 44.7\%                                       & 1{,}899                                 & 2.4            & 18.4M                                     & 136                                      & 297                                      & 20.9m \\
                 & \codex         & \openailogo~GPT-5.4       & 26              & 25.0\%                                       & 104           & 0                & 25.0\%                                       & 1{,}595                                 & 1.6            & 19.9M                                     & 164                                      & 344                                      & 19.4m \\
                 & \claude        & \claudelogo~Opus 4.6      & 29              & 27.9\%                                       & 43            & 61               & \textbf{58.1}\%                        & 7{,}914                                 & 0.4            & 40.8M                                     & 537                                      & 718                                      & 68.8m \\
                 & \ocode         & \zailogo~GLM-5            & 6               & 5.8\%                                        & 96            & 8                & 6.2\%                                        & 2{,}068                                 & 0.3            & 9.6M                                      & 105                                      & 140                                      & 47.1m \\
                 & \ocode         & \kimilogo~Kimi K2.5       & 3               & 2.9\%                                        & 103           & 1                & 2.9\%                                        & 1{,}547                                 & 0.2            & 8.6M                                      & 91                                       & 103                                      & 20.8m \\
                 & \ocode         & \minimaxlogo~MiniMax M2.5 & 0               & 0.0\%                                        & 100           & 4                & 0.0\%                                        & 1{,}742                                 & 0.0            & 6.2M                                      & 94                                       & 94                                       & 31.8m \\
  \midrule
  \multirow{6}{*}{\rotatebox[origin=c]{90}{Linux}}
                 & \codex         & \openailogo~GPT-5.5       & 106       & \textbf{77.4}\%                        & 125     & 12         & 83.2\%                                 & 1{,}219                           & 8.7      & 11.3M                                     & 86                                 & 173                                      & 29.3m \\
                 & \codex         & \openailogo~GPT-5.4       & 72        & 52.6\%                                 & 124     & 13         & 56.5\%                                 & 1{,}306                           & 5.5      & 14.2M                               & 115                                & 178                                & 43.6m \\
                 & \claude        & \claudelogo~Opus 4.6      & 54        & 39.4\%                                 & 57      & 80         & \textbf{91.2}\%                        & 905                               & 6.0      & 21.6M                               & 310                                & 381                                & 67.9m \\
                 & \ocode         & \zailogo~GLM-5            & 5         & 3.6\%                                  & 134     & 3          & 3.7\%                                  & 1{,}254                           & 0.4      & 6.4M                                & 86                                 & 101                                & 36.7m \\
                 & \ocode         & \kimilogo~Kimi K2.5       & 3         & 2.2\%                                  & 137     & 0          & 2.2\%                                  & 1{,}006                           & 0.3      & 6.6M                                & 83                                 & 85                                 & 21.6m \\
                 & \ocode         & \minimaxlogo~MiniMax M2.5 & 2               & 1.5\%                                        & 132     & 5          & 1.5\%                                        & 1{,}371                           & 0.1      & 5.7M                                & 88                                 & 87                                 & 45.1m \\
  \bottomrule
\end{tabular}

  }
\end{table*}

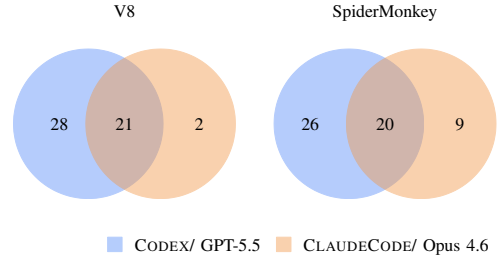
\begin{figure}[t]
  \centering
  \resizebox{0.8\linewidth}{!}{\begin{tikzpicture}[gnuplot]
\tikzset{every node/.append style={font={\fontsize{8.0pt}{9.6pt}\selectfont}}}
\path (0.000,0.000) rectangle (8.382,4.445);
  \gpfill{rgb color={0.357,0.561,0.976},opacity=0.45} (2.734,2.222)--(2.732,2.284)--(2.728,2.346)--(2.719,2.407)%
    --(2.708,2.468)--(2.694,2.529)--(2.676,2.588)--(2.655,2.647)--(2.631,2.705)%
    --(2.605,2.761)--(2.575,2.815)--(2.543,2.868)--(2.507,2.920)--(2.469,2.969)%
    --(2.429,3.016)--(2.386,3.061)--(2.341,3.104)--(2.294,3.144)--(2.245,3.182)%
    --(2.193,3.218)--(2.140,3.250)--(2.086,3.280)--(2.030,3.306)--(1.972,3.330)%
    --(1.913,3.351)--(1.854,3.369)--(1.793,3.383)--(1.732,3.394)--(1.671,3.403)%
    --(1.609,3.407)--(1.547,3.409)--(1.484,3.407)--(1.422,3.403)--(1.361,3.394)%
    --(1.300,3.383)--(1.239,3.369)--(1.180,3.351)--(1.121,3.330)--(1.063,3.306)%
    --(1.007,3.280)--(0.953,3.250)--(0.900,3.218)--(0.848,3.182)--(0.799,3.144)%
    --(0.752,3.104)--(0.707,3.061)--(0.664,3.016)--(0.624,2.969)--(0.586,2.920)%
    --(0.550,2.868)--(0.518,2.815)--(0.488,2.761)--(0.462,2.705)--(0.438,2.647)%
    --(0.417,2.588)--(0.399,2.529)--(0.385,2.468)--(0.374,2.407)--(0.365,2.346)%
    --(0.361,2.284)--(0.359,2.222)--(0.361,2.159)--(0.365,2.097)--(0.374,2.036)%
    --(0.385,1.975)--(0.399,1.914)--(0.417,1.855)--(0.438,1.796)--(0.462,1.738)%
    --(0.488,1.682)--(0.518,1.628)--(0.550,1.575)--(0.586,1.523)--(0.624,1.474)%
    --(0.664,1.427)--(0.707,1.382)--(0.752,1.339)--(0.799,1.299)--(0.848,1.261)%
    --(0.900,1.225)--(0.953,1.193)--(1.007,1.163)--(1.063,1.137)--(1.121,1.113)%
    --(1.180,1.092)--(1.239,1.074)--(1.300,1.060)--(1.361,1.049)--(1.422,1.040)%
    --(1.484,1.036)--(1.546,1.034)--(1.609,1.036)--(1.671,1.040)--(1.732,1.049)%
    --(1.793,1.060)--(1.854,1.074)--(1.913,1.092)--(1.972,1.113)--(2.030,1.137)%
    --(2.086,1.163)--(2.140,1.193)--(2.193,1.225)--(2.245,1.261)--(2.294,1.299)%
    --(2.341,1.339)--(2.386,1.382)--(2.429,1.427)--(2.469,1.474)--(2.507,1.523)%
    --(2.543,1.575)--(2.575,1.628)--(2.605,1.682)--(2.631,1.738)--(2.655,1.796)%
    --(2.676,1.855)--(2.694,1.914)--(2.708,1.975)--(2.719,2.036)--(2.728,2.097)--(2.732,2.159)--cycle;
  \gpfill{rgb color={0.949,0.600,0.290},opacity=0.45} (3.883,2.222)--(3.881,2.284)--(3.877,2.346)--(3.868,2.407)%
    --(3.857,2.468)--(3.843,2.529)--(3.825,2.588)--(3.804,2.647)--(3.780,2.705)%
    --(3.754,2.761)--(3.724,2.815)--(3.692,2.868)--(3.656,2.920)--(3.618,2.969)%
    --(3.578,3.016)--(3.535,3.061)--(3.490,3.104)--(3.443,3.144)--(3.394,3.182)%
    --(3.342,3.218)--(3.289,3.250)--(3.235,3.280)--(3.179,3.306)--(3.121,3.330)%
    --(3.062,3.351)--(3.003,3.369)--(2.942,3.383)--(2.881,3.394)--(2.820,3.403)%
    --(2.758,3.407)--(2.696,3.409)--(2.633,3.407)--(2.571,3.403)--(2.510,3.394)%
    --(2.449,3.383)--(2.388,3.369)--(2.329,3.351)--(2.270,3.330)--(2.212,3.306)%
    --(2.156,3.280)--(2.102,3.250)--(2.049,3.218)--(1.997,3.182)--(1.948,3.144)%
    --(1.901,3.104)--(1.856,3.061)--(1.813,3.016)--(1.773,2.969)--(1.735,2.920)%
    --(1.699,2.868)--(1.667,2.815)--(1.637,2.761)--(1.611,2.705)--(1.587,2.647)%
    --(1.566,2.588)--(1.548,2.529)--(1.534,2.468)--(1.523,2.407)--(1.514,2.346)%
    --(1.510,2.284)--(1.508,2.222)--(1.510,2.159)--(1.514,2.097)--(1.523,2.036)%
    --(1.534,1.975)--(1.548,1.914)--(1.566,1.855)--(1.587,1.796)--(1.611,1.738)%
    --(1.637,1.682)--(1.667,1.628)--(1.699,1.575)--(1.735,1.523)--(1.773,1.474)%
    --(1.813,1.427)--(1.856,1.382)--(1.901,1.339)--(1.948,1.299)--(1.997,1.261)%
    --(2.049,1.225)--(2.102,1.193)--(2.156,1.163)--(2.212,1.137)--(2.270,1.113)%
    --(2.329,1.092)--(2.388,1.074)--(2.449,1.060)--(2.510,1.049)--(2.571,1.040)%
    --(2.633,1.036)--(2.696,1.034)--(2.758,1.036)--(2.820,1.040)--(2.881,1.049)%
    --(2.942,1.060)--(3.003,1.074)--(3.062,1.092)--(3.121,1.113)--(3.179,1.137)%
    --(3.235,1.163)--(3.289,1.193)--(3.342,1.225)--(3.394,1.261)--(3.443,1.299)%
    --(3.490,1.339)--(3.535,1.382)--(3.578,1.427)--(3.618,1.474)--(3.656,1.523)%
    --(3.692,1.575)--(3.724,1.628)--(3.754,1.682)--(3.780,1.738)--(3.804,1.796)%
    --(3.825,1.855)--(3.843,1.914)--(3.857,1.975)--(3.868,2.036)--(3.877,2.097)--(3.881,2.159)--cycle;
  \gpfill{rgb color={0.357,0.561,0.976},opacity=0.45} (6.872,2.222)--(6.870,2.284)--(6.866,2.346)--(6.857,2.407)%
    --(6.846,2.468)--(6.832,2.529)--(6.814,2.588)--(6.793,2.647)--(6.769,2.705)%
    --(6.743,2.761)--(6.713,2.815)--(6.681,2.868)--(6.645,2.920)--(6.607,2.969)%
    --(6.567,3.016)--(6.524,3.061)--(6.479,3.104)--(6.432,3.144)--(6.383,3.182)%
    --(6.331,3.218)--(6.278,3.250)--(6.224,3.280)--(6.168,3.306)--(6.110,3.330)%
    --(6.051,3.351)--(5.992,3.369)--(5.931,3.383)--(5.870,3.394)--(5.809,3.403)%
    --(5.747,3.407)--(5.685,3.409)--(5.622,3.407)--(5.560,3.403)--(5.499,3.394)%
    --(5.438,3.383)--(5.377,3.369)--(5.318,3.351)--(5.259,3.330)--(5.201,3.306)%
    --(5.145,3.280)--(5.091,3.250)--(5.038,3.218)--(4.986,3.182)--(4.937,3.144)%
    --(4.890,3.104)--(4.845,3.061)--(4.802,3.016)--(4.762,2.969)--(4.724,2.920)%
    --(4.688,2.868)--(4.656,2.815)--(4.626,2.761)--(4.600,2.705)--(4.576,2.647)%
    --(4.555,2.588)--(4.537,2.529)--(4.523,2.468)--(4.512,2.407)--(4.503,2.346)%
    --(4.499,2.284)--(4.497,2.222)--(4.499,2.159)--(4.503,2.097)--(4.512,2.036)%
    --(4.523,1.975)--(4.537,1.914)--(4.555,1.855)--(4.576,1.796)--(4.600,1.738)%
    --(4.626,1.682)--(4.656,1.628)--(4.688,1.575)--(4.724,1.523)--(4.762,1.474)%
    --(4.802,1.427)--(4.845,1.382)--(4.890,1.339)--(4.937,1.299)--(4.986,1.261)%
    --(5.038,1.225)--(5.091,1.193)--(5.145,1.163)--(5.201,1.137)--(5.259,1.113)%
    --(5.318,1.092)--(5.377,1.074)--(5.438,1.060)--(5.499,1.049)--(5.560,1.040)%
    --(5.622,1.036)--(5.685,1.034)--(5.747,1.036)--(5.809,1.040)--(5.870,1.049)%
    --(5.931,1.060)--(5.992,1.074)--(6.051,1.092)--(6.110,1.113)--(6.168,1.137)%
    --(6.224,1.163)--(6.278,1.193)--(6.331,1.225)--(6.383,1.261)--(6.432,1.299)%
    --(6.479,1.339)--(6.524,1.382)--(6.567,1.427)--(6.607,1.474)--(6.645,1.523)%
    --(6.681,1.575)--(6.713,1.628)--(6.743,1.682)--(6.769,1.738)--(6.793,1.796)%
    --(6.814,1.855)--(6.832,1.914)--(6.846,1.975)--(6.857,2.036)--(6.866,2.097)--(6.870,2.159)--cycle;
  \gpfill{rgb color={0.949,0.600,0.290},opacity=0.45} (8.021,2.222)--(8.019,2.284)--(8.015,2.346)--(8.006,2.407)%
    --(7.995,2.468)--(7.981,2.529)--(7.963,2.588)--(7.942,2.647)--(7.918,2.705)%
    --(7.892,2.761)--(7.862,2.815)--(7.830,2.868)--(7.794,2.920)--(7.756,2.969)%
    --(7.716,3.016)--(7.673,3.061)--(7.628,3.104)--(7.581,3.144)--(7.532,3.182)%
    --(7.480,3.218)--(7.427,3.250)--(7.373,3.280)--(7.317,3.306)--(7.259,3.330)%
    --(7.200,3.351)--(7.141,3.369)--(7.080,3.383)--(7.019,3.394)--(6.958,3.403)%
    --(6.896,3.407)--(6.834,3.409)--(6.771,3.407)--(6.709,3.403)--(6.648,3.394)%
    --(6.587,3.383)--(6.526,3.369)--(6.467,3.351)--(6.408,3.330)--(6.350,3.306)%
    --(6.294,3.280)--(6.240,3.250)--(6.187,3.218)--(6.135,3.182)--(6.086,3.144)%
    --(6.039,3.104)--(5.994,3.061)--(5.951,3.016)--(5.911,2.969)--(5.873,2.920)%
    --(5.837,2.868)--(5.805,2.815)--(5.775,2.761)--(5.749,2.705)--(5.725,2.647)%
    --(5.704,2.588)--(5.686,2.529)--(5.672,2.468)--(5.661,2.407)--(5.652,2.346)%
    --(5.648,2.284)--(5.646,2.222)--(5.648,2.159)--(5.652,2.097)--(5.661,2.036)%
    --(5.672,1.975)--(5.686,1.914)--(5.704,1.855)--(5.725,1.796)--(5.749,1.738)%
    --(5.775,1.682)--(5.805,1.628)--(5.837,1.575)--(5.873,1.523)--(5.911,1.474)%
    --(5.951,1.427)--(5.994,1.382)--(6.039,1.339)--(6.086,1.299)--(6.135,1.261)%
    --(6.187,1.225)--(6.240,1.193)--(6.294,1.163)--(6.350,1.137)--(6.408,1.113)%
    --(6.467,1.092)--(6.526,1.074)--(6.587,1.060)--(6.648,1.049)--(6.709,1.040)%
    --(6.771,1.036)--(6.834,1.034)--(6.896,1.036)--(6.958,1.040)--(7.019,1.049)%
    --(7.080,1.060)--(7.141,1.074)--(7.200,1.092)--(7.259,1.113)--(7.317,1.137)%
    --(7.373,1.163)--(7.427,1.193)--(7.480,1.225)--(7.532,1.261)--(7.581,1.299)%
    --(7.628,1.339)--(7.673,1.382)--(7.716,1.427)--(7.756,1.474)--(7.794,1.523)%
    --(7.830,1.575)--(7.862,1.628)--(7.892,1.682)--(7.918,1.738)--(7.942,1.796)%
    --(7.963,1.855)--(7.981,1.914)--(7.995,1.975)--(8.006,2.036)--(8.015,2.097)--(8.019,2.159)--cycle;
\gpfill{rgb color={0.357,0.561,0.976},color=.!45} (1.854,0.192)--(2.083,0.192)--(2.083,0.421)--(1.854,0.421)--cycle;
\gpfill{rgb color={0.949,0.600,0.290},color=.!45} (4.535,0.192)--(4.765,0.192)--(4.765,0.421)--(4.535,0.421)--cycle;
\gpcolor{color=gp lt color border}
\node[gp node center] at (1.087,2.222) {28};
\node[gp node center] at (2.122,2.222) {21};
\node[gp node center] at (3.309,2.222) {2};
\node[gp node center] at (5.072,2.222) {26};
\node[gp node center] at (6.259,2.222) {20};
\node[gp node center] at (7.447,2.222) {9};
\node[gp node center] at (2.122,4.023) {V8};
\node[gp node center] at (6.259,4.023) {SpiderMonkey};
\node[gp node left] at (2.160,0.306) {\codex / GPT-5.5};
\node[gp node left] at (4.842,0.306) {\claude / Opus 4.6};
\gpdefrectangularnode{gp plot 1}{\pgfpoint{0.053cm}{0.000cm}}{\pgfpoint{8.328cm}{4.444cm}}
\end{tikzpicture}
}
  \caption{Verified-instance overlap between \codex/GPT-5.5 and \claude/Opus 4.6.
  The two agents share a common core but each solves instances the other misses.}
  \label{fig:config-overlap}
\end{figure}

\PP{Overall performance}
\autoref{tab:rq1-main} reports per-instance success for all configurations.
The strongest configuration, \codex GPT-5.5, solves 201 of 344 instances (58.4\%), demonstrating the remarkable progress of frontier models on long-horizon security tasks.
In fact, all studied closed-weight agents achieve over a 20\% solved rate, while open-weight models stay in single digits on every target.
\codex and \claude solve different instances, as \autoref{fig:config-overlap} shows.
On SpiderMonkey, \codex GPT-5.5 and \claude Opus 4.6 agree on only 20 instances.
Each solves instances the other misses, lifting their union to 55 against 46 for the better of the two.
The same divergence holds on V8, where the union reaches 51 against 49.
It shrinks on Linux, where the stronger configuration almost always subsumes the other's solves.
This phenomenon echoes ProgramBench's~\citep{yang2026programbench} observation that model-specific coding habits dominate which tasks get solved.

\PP{Target reachability gates headline capability}
Overall, the agents tend to solve more Linux instances than the browser engines, with the frontier configuration (\codex GPT-5.5) reaching 77.4\% on Linux but only 47.6\% on V8 and 44.2\% on SpiderMonkey.
These families expose different input-to-bug paths.
A kernel bug is reached through an explicit and structured syscall interface, so the search is over which syscalls to issue and in what order, and fuzzers already navigate this space~\citep{pailoor2018moonshine,sun2021healer}.
A browser engine includes a compiler pipeline, so a JavaScript PoC must indirectly reach the bug by composing language features that drive the engine through optimization tiers into a specific JIT, garbage-collection, or object-layout state.
These subtly conflicting optimization assumptions make engine bugs hard to trigger~\citep{wachter2025dumpling,han2019codealchemist}.
Model progress tracks this reachability gap: from GPT-5.4 to GPT-5.5, \codex adds 33 instances across the two browser engines (V8 36 to 49, SpiderMonkey 26 to 46) and 34 on Linux (72 to 106), so the largest single-family gain lands on the most reachable target.
A benchmark built on either family alone would therefore misstate agent capability and its trajectory.

\PP{Timeouts drive Claude's low headline rank}
The \emph{Compl. Sol. Rate} column, the solve rate over instances that finish within budget, separates the score lost to unfinished runs from the score lost to wrong PoCs.
For \codex GPT-5.5 it barely moves on the browser engines, from 47.6\% to 48.0\% on V8 where it times out on a single instance, and rises from 77.4\% to 83.2\% on Linux where 12 of its runs exhaust the budget.
\claude shows the opposite pattern, rising from 22.3\% to 61.1\% on V8, 27.9\% to 58.1\% on SpiderMonkey, and 39.4\% to 91.2\% on Linux.
Among completed instances, it leads on all three targets, despite ranking behind the frontier configuration on every target by overall rate.
We nonetheless keep \emph{Overall Sol. Rate} as the headline, since every configuration receives the same 90-minute budget and \claude spends it most heavily, averaging 36 to 41 million tokens and 650 to 720 tool calls per browser instance.
Finishing within that budget is part of the task rather than a resource handicap.

\subsection{RQ2: Grading Effects}
\label{s:attribution-pipeline}
\begin{table}[t]
  \centering
  \caption{Solved-instance counts under two rule-based graders (\Jv, \Jvf) and our three-image judge (\Jllm), each reporting its own automatic output with \emph{FP} and \emph{FN} measured against the adjudicated ground truth.
    Ground truth is the manually adjudicated label of every candidate PoC: we inspect all verified verdicts and every illegal verdict that crashes the vulnerable image, the only cases that can hide a missed valid PoC.
    \Jv only over-credits off-target failures (\emph{FP}), while \Jvf additionally under-credits valid PoCs (\emph{FN}).
\Jllm reports its raw verdict, so its residual \emph{FP}/\emph{FN} are the disagreements manual review later resolves.
  The reported \Jvf requires a clean fixed-image exit.}
  \label{tab:rq2-attribution}
  \resizebox{\columnwidth}{!}{
    \setlength{\tabcolsep}{3.2pt}
\ra{1.1}
\begin{tabular}{@{}c l l r r | r r r | r r r@{}}
  \toprule
                 &                &                           & \multicolumn{2}{c}{\Jv} & \multicolumn{3}{c}{\Jvf} & \multicolumn{3}{c}{\Jllm (Ours)} \\
  \cmidrule(lr){4-5}\cmidrule(lr){6-8}\cmidrule(lr){9-11}
  \textbf{Proj.} & \textbf{Agent} & \textbf{Model}            & \textbf{Sol.}                 & \textbf{FP}                    & \textbf{Sol.} & \textbf{FN} & \textbf{FP} & \textbf{Sol.} & \textbf{FN} & \textbf{FP} \\
  \midrule
  \multirow{6}{*}{\rotatebox[origin=c]{90}{V8}}
                 & \codex         & \openailogo~GPT-5.5       & 58                            & \O9                            & 18            & 31          & \O0         & 49          & \O0   & \O0 \\
                 & \codex         & \openailogo~GPT-5.4       & 42                            & \O6                            & 16            & 21          & \O1         & 36          & \O0   & \O0 \\
                 & \claude        & \claudelogo~Opus 4.6      & 42                      & 19                       & 17      & 12    & \O6  & 23          & \O0   & \O0 \\
                 & \ocode         & \zailogo~GLM-5            & \O4                           & \O2                            & \O2           & \O1         & \O1         & \O2         & \O0   & \O0 \\
                 & \ocode         & \kimilogo~Kimi K2.5       & \O7                           & \O5                            & \O0           & \O2         & \O0         & \O2         & \O0   & \O0 \\
                 & \ocode         & \minimaxlogo~MiniMax M2.5 & \O2                           & \O2                            & \O0           & \O0         & \O0         & \O0         & \O0   & \O0 \\
  \midrule
  \multirow{6}{*}{\rotatebox[origin=c]{90}{SpiderMonkey}}
                 & \codex         & \openailogo~GPT-5.5       & 68                            & 22                             & 40            & 12          & \O6         & 46          & \O0   & \O0 \\
                 & \codex         & \openailogo~GPT-5.4       & 34                            & \O8                            & 20            & \O8         & \O2         & 26          & \O0   & \O0 \\
                 & \claude        & \claudelogo~Opus 4.6      & 34                            & \O5                            & 22            & \O8         & \O1         & 29          & \O0   & \O0 \\
                 & \ocode         & \zailogo~GLM-5            & \O7                           & \O1                            & \O6           & \O0         & \O0         & \O6         & \O0   & \O0 \\
                 & \ocode         & \kimilogo~Kimi K2.5       & \O6                           & \O3                            & \O4           & \O1         & \O2         & \O3         & \O0   & \O0 \\
                 & \ocode         & \minimaxlogo~MiniMax M2.5 & \O2                           & \O2                            & \O1           & \O0         & \O1         & \O0         & \O0   & \O0 \\
  \midrule
  \multirow{6}{*}{\rotatebox[origin=c]{90}{Linux}}
                 & \codex         & \openailogo~GPT-5.5       & 116                     & 10                       & 52      & 56    & \O2   & 102   & \O7   & \O3 \\
                 & \codex         & \openailogo~GPT-5.4       & 84                      & 12                       & 44      & 32    & \O4   & 67    & \O6   & \O1 \\
                 & \claude        & \claudelogo~Opus 4.6      & 64                      & 10                       & 37      & 20    & \O3   & 54          & \O0   & \O0 \\
                 & \ocode         & \zailogo~GLM-5            & \O5                     & \O0                      & \O5     & \O0   & \O0         & \O5         & \O0   & \O0 \\
                 & \ocode         & \kimilogo~Kimi K2.5       & \O4                     & \O1                      & \O3     & \O0   & \O0         & \O3         & \O0   & \O0 \\
                 & \ocode         & \minimaxlogo~MiniMax M2.5 & \O2                     & \O0                      & \O2     & \O0   & \O0         & \O2         & \O0   & \O0 \\
  \bottomrule
\end{tabular}

  }
\end{table}

We then compare our three-image judge \Jllm with two rule-based graders (\Jv and \Jvf) and show results in \autoref{tab:rq2-attribution}, where each grader reports its own automatic output.
\Jv credits any vulnerable-image crash or sanitizer report, while \Jvf credits a vulnerable-image failure that the fixed image no longer reproduces.

\noindent{\bf{\Jv overstates capability.}}
\Jv over-credits off-target failures because a PoC may crash through an unrelated bug.
On V8, \codex GPT-5.5 triggers 58 instances but only 49 are attributed to the target, and \claude triggers 42 but verifies 23, so \Jv can naively inflate the weaker configuration more than the stronger one.
The main reason is that weaker configurations may fail to precisely follow the detailed instructions and tend to generate more irrelevant PoCs.
The effect is even larger for open-weight models, whose speculative PoCs frequently crash but rarely on the reported bug.
For example, \ocode Kimi-K2.5 crashes 7 V8 instances for 2 verified, so \Jv more than triples its score.

\noindent{\bf{\Jvf discards valid PoCs.}} While \Jvf can reduce the false positive issue of \Jv through two-image validation, it incurs additional false negatives.
The strict rule discards a final success whenever none of its target-aligned PoCs both crashes the vulnerable image and produces the rule's exact clean exit on the fixed image.
In fact, \Jvf discards 204 of the 464 final verified run-instance successes.
This performance loss concentrates in the strongest agents, removing 56 of \codex GPT-5.5's 106 Linux instances and 31 of its 49 V8 instances, so \Jvf understates the frontier more than the open-weight models and again reorders the field.

These false negatives arise when the fixed run ends with a benign nonzero exit, stops before completing because of an infrastructure failure, or reproduces a target-aligned crash after the historical patch.
The exact-exit rule rejects all three cases by construction.
\Jllm instead attributes the vulnerable-image crash to the target source and error type, then uses the fixed and latest executions to identify evidence that contradicts this attribution.
It does not require an exact clean exit from the fixed image and accepts a persistent crash only when the evidence remains target-aligned.
Reading fixed-image logs relaxes \Jvf but cannot establish target attribution.

\noindent{\bf{The performance of \Jllm.}}
On Linux, \Jv adds 10 false positives for GPT-5.5 (116 triggers versus 106 verified).
These cases satisfy the crash-only rule but fail target attribution.
The strict \Jvf reports 52 solves, with 50 final successes and two false positives, but misses 56 other final successes.
\Jllm reports its raw verdict for the same run, crediting 102 solves against the 106 ground-truth successes with three false positives and seven false negatives.
Its seven false negatives are a strict subset of the 56 that \Jvf discards, so \Jllm recovers 49 of those correct PoCs without introducing any false negative that \Jvf would have caught.
The two graders instead make their false positives on disjoint instances for opposite reasons: \Jvf credits two off-target crashes whose fixed image happens to run clean, both of which \Jllm rejects on attribution, while \Jllm's three arise from crashes that persist on a non-clean fixed image at a target-adjacent site, all of which \Jvf rejects on its fixed-clean rule.
Across all runs, \Jllm credits 455 of the 464 solved run-instances with four false positives and 13 false negatives, a 99.1\% precision and 97.2\% recall against ground truth, and every residual error is a Linux \codex verdict that our manual adjudication corrects, the same review that fixes the ground-truth labels.
Both rule-based graders carry an order of magnitude more error on the same runs.
\Jllm makes no error on either browser engine because symbolized ASAN and DCHECK reports pin the crash to an exact file and line, so target attribution is unambiguous, whereas Linux KASAN evidence exposes the residual failure modes: infrastructure gaps in an image execution and sanitizer classes that must match a related but distinct crash type.

The recovered Linux PoCs carry target-specific evidence.
CVE-2023-31436, for example, fires the expected \cc{KASAN\_OOB} at \cc{qfq\_activate\_agg} in the target file \cc{net/sched/sch\_qfq.c} on the vulnerable image, while the latest image rejects the input cleanly.
The fixed image emits the same KASAN class at the same function.
\Jv credits the crash without establishing attribution, while \Jvf rejects it because the fixed-image exit is not clean.
This case shows why grading must assess target attribution rather than require a clean fixed-image exit.

Across 3{,}957 judge-eligible candidate PoCs, \Jllm returns a definitive verdict for 3{,}934 (99.4\%), returns \cc{unsure} for 22, and encounters one service error.
All 22 \cc{unsure} outcomes are Linux candidates whose fixed or latest evidence contains an infrastructure gap or a residual sibling crash.
Manual adjudication resolves 20 as \cc{verified} and two as \cc{illegal}, and labels the service-error candidate \cc{illegal} from its clean executions on all three images.
This 90.9\% confirmation rate further justifies \sys's default configuration of treating both \cc{unsure} and \cc{verified} cases as successful attempts.

To establish ground truth for the above analysis, we manually inspect every automatically \cc{verified} verdict and every automatically \cc{illegal} verdict that crashes the vulnerable image.
The second set matters because a false negative requires a valid PoC that the judge discarded, and a PoC that never crashes the vulnerable image cannot reproduce a memory-safety bug.
Only 132 of the 3{,}486 \cc{illegal} verdicts produce a genuine crash, so the remaining 3{,}354 are illegal by construction and the inspected set covers every candidate that could hide a missed valid PoC.
The inspection confirms 447 of the 449 \cc{verified} verdicts, a 99.6\% precision for \cc{verified} verdicts, and corrects the other two to \cc{illegal}.
Among the 132 crashing \cc{illegal} verdicts, 119 are confirmed \cc{illegal} because the crash lands outside the target source, carries the wrong sanitizer class, or persists identically on the fixed image, while 13 recover a valid target-aligned PoC and become \cc{verified}.
Counting both \cc{verified} and \cc{unsure} verdicts as success predictions, the judge reaches 99.2\% precision at the PoC level (467 of 471) and 99.1\% at the run-instance level (451 of 455).
Together with the 22 \cc{unsure} resolutions, these corrections yield final totals of 469 \cc{verified}, 3{,}486 \cc{illegal}, and 28 pre-screened \cc{invalid} candidate PoCs, corresponding to 464 solved run-instances that form the reference outcomes in \autoref{tab:rq2-attribution}.

Lastly, since existing benchmarks~\citep{wang2025cybergym,lee2025sec,wang2025vulnrepaireval} typically apply \Jvf, we additionally test a stronger rule-based extension that further uses the latest image to recover PoCs rejected when the fixed execution is not clean.
The extension credits a candidate PoC when the vulnerable execution satisfies the crash criterion used by \Jv and either the fixed or the latest execution is clean.
It credits an evaluated agent run when at least one candidate PoC satisfies these conditions.
Across all evaluated runs, this stronger extension still produces 61 run-level false positives relative to the adjudicated reference outcomes: 16 on V8, 20 on SpiderMonkey, and 25 on Linux.
The result shows that differential execution across all three images does not establish target attribution: the vulnerable-image crash can still originate from the wrong source or exhibit the wrong error type for the target vulnerability, demonstrating the necessity and effectiveness of our \Jllm.

\begin{figure}[t]
  \centering
  \includegraphics[width=0.95\columnwidth]{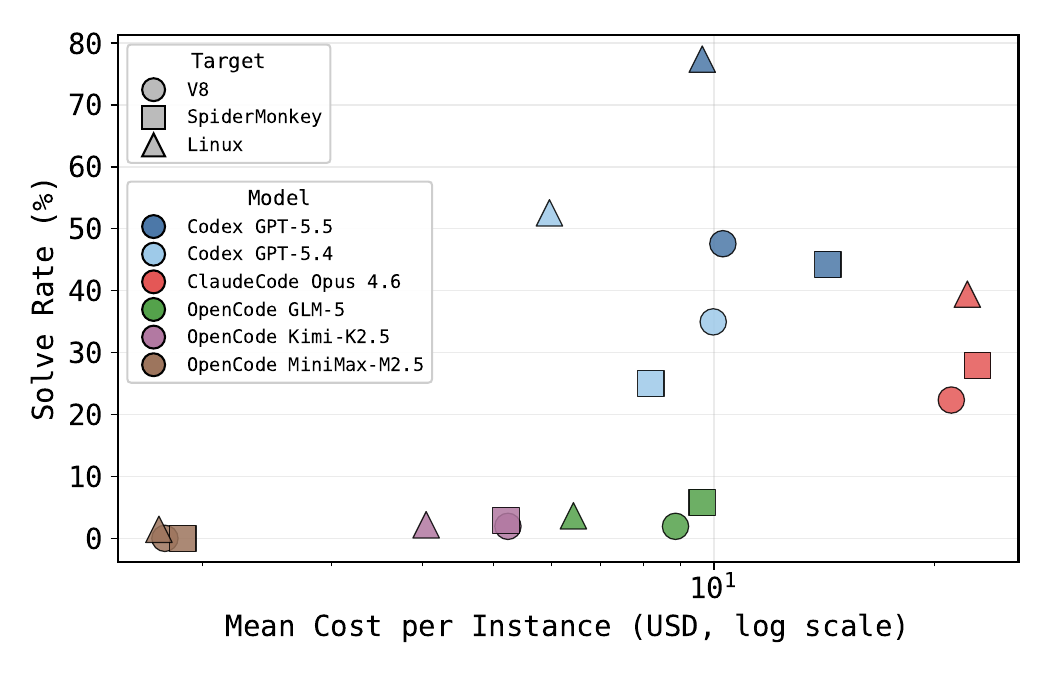}
  \caption{Per-instance cost (x-axis) versus solve rate (y-axis) across all six configurations and three targets.
    Each marker is one (model, target) pair, with color encoding the model and shape encoding the target.
  }
  \label{fig:cost-solve}
\end{figure}

\begin{figure}[t]
  \centering
  \includegraphics[width=\columnwidth]{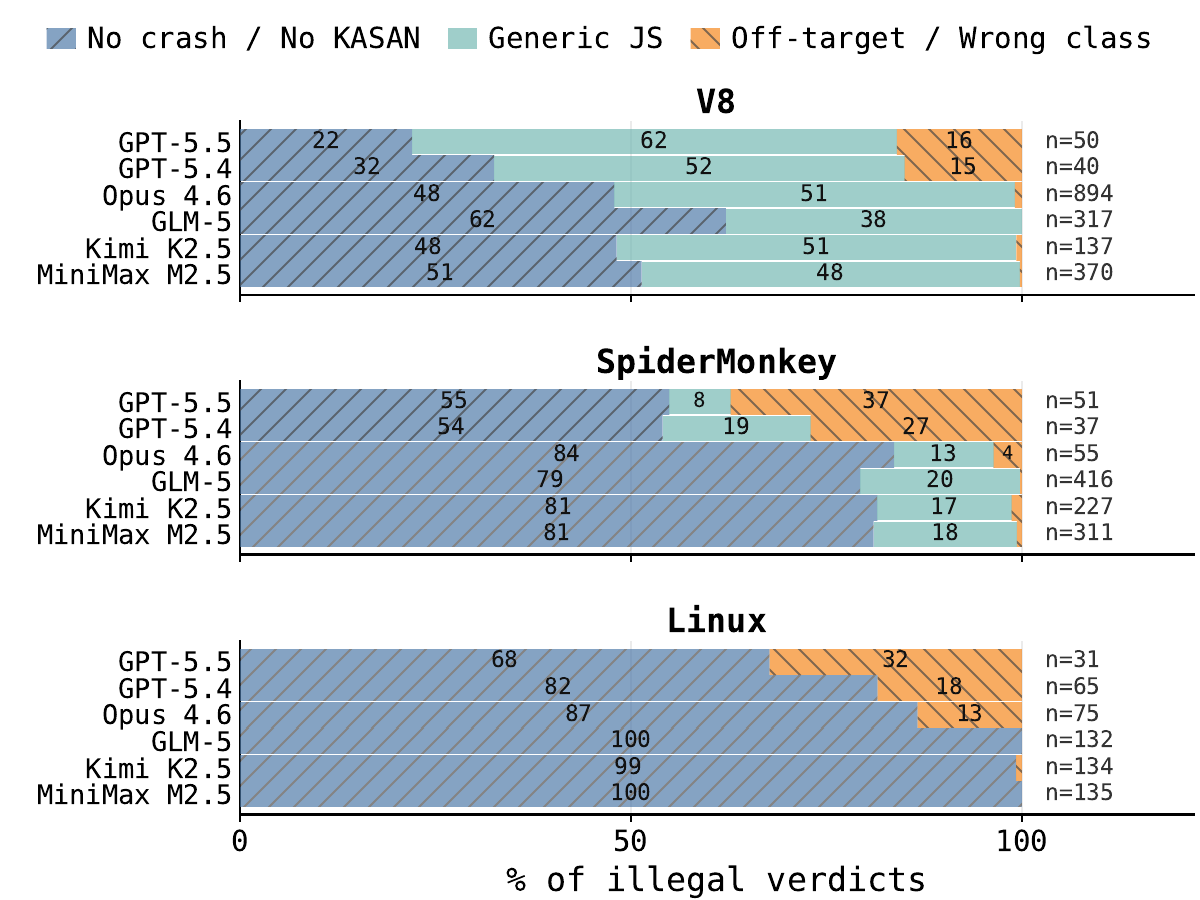}
  \caption{Per-PoC failure modes as a share of each run's \cc{illegal} verdicts (\emph{n} is the illegal-verdict count).
  \emph{No crash}/\emph{No KASAN} produce no crash signal, \emph{Generic JS} aborts with a benign language-level exception, and \emph{Off-target}/\emph{Wrong class} crash at the wrong site or vulnerability type.}
  \label{fig:failure-modes}
\end{figure}

\begin{figure*}[t]
  \centering
  \includegraphics[width=0.95\textwidth]{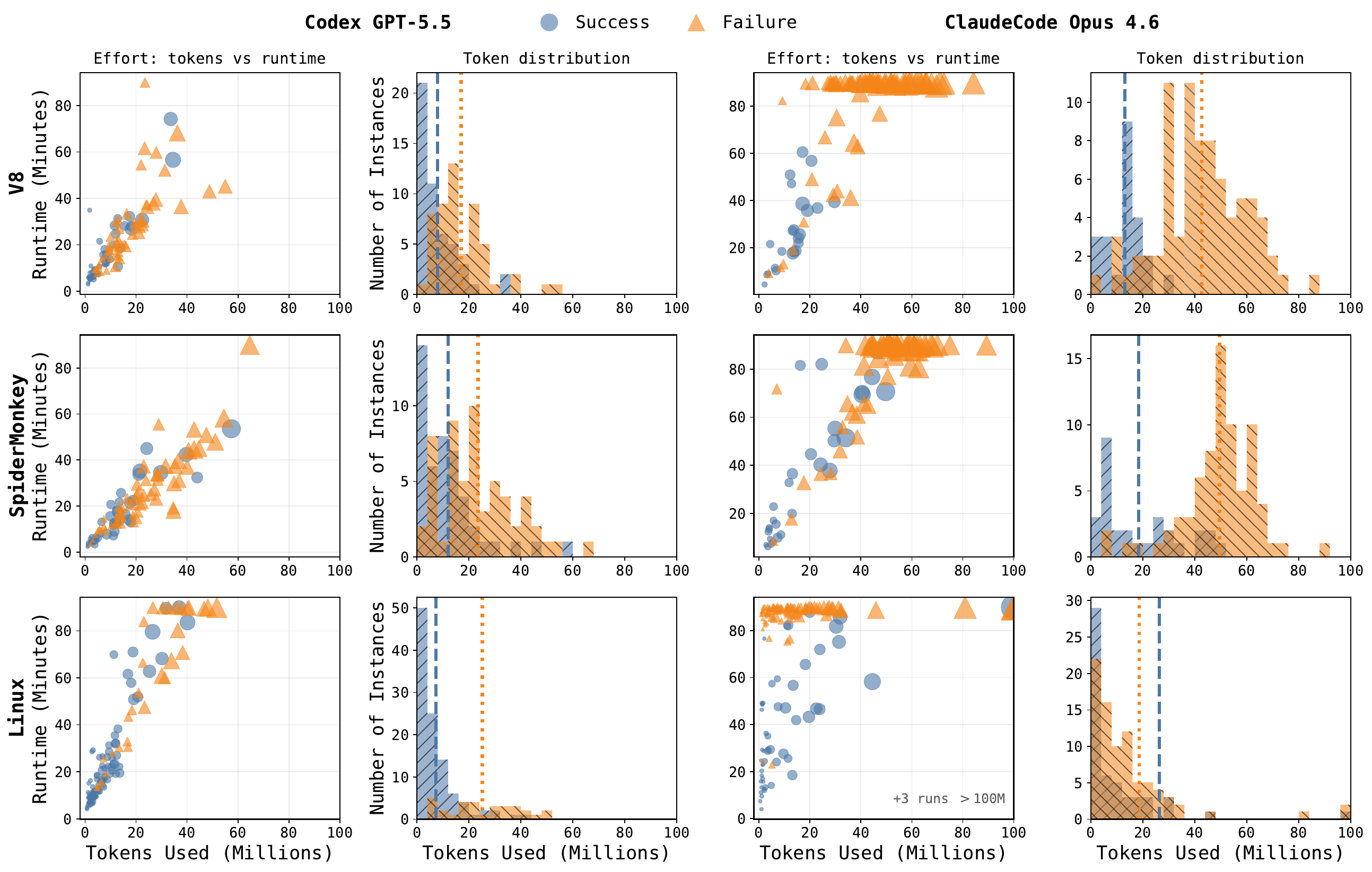}
  \caption{Per-instance effort by success outcome for \codex GPT-5.5 and \claude Opus 4.6 across the three targets (rows), omitting open-weight runs that verify too few instances.
  Per configuration, the left panel plots tokens against runtime (marker area proportional to tool calls) and the right shows the token distribution with dashed (success) and dotted (failure) means.}
  \label{fig:effort-tokens}
\end{figure*}

\subsection{RQ3: Further Performance Analysis}
\label{s:deep-analysis}

\PP{Failure runs often consume more tokens}
Failed runs consume more tokens and more wall-clock time than successful ones in every panel of \autoref{fig:effort-tokens}.
The gap holds even for the frontier \codex GPT-5.5, whose failures consume over 2$\times$ the tokens of its successes.
Runtime moves with tokens: \claude's failed and successful runs on average take over 80 minutes and less than 40 minutes, respectively.
As shown in the third column of \autoref{fig:effort-tokens}, the failures cluster below the wall-clock ceiling.
This pattern is consistent with reports that context length degrades model performance beyond a task-specific threshold~\citep{du2025context} and with the SWE-Bench Pro finding that non-submitting runs concentrate in long-context and stuck-in-loop categories~\citep{deng2025swebenchpro}.

\PP{Higher cost does not buy capability}
\autoref{fig:cost-solve} shows that cost and capability are decoupled: \claude (red) sits at the highest per-instance cost yet never the highest solve rate.
Its aggregate per-solved-instance cost is 3.7$\times$ that of \codex GPT-5.5 (blue, \$71.58 versus \$19.23).
Across targets, \claude costs 2.5--4.5$\times$ as much per success.
The open-weight models are cheapest per instance but most expensive per success, with GLM-5 exceeding \$450 per solved V8 instance due to its low solve rate.
Cost efficiency follows attributable capability rather than raw token expenditure.

\PP{Open-weight models favor shallow exploration}
The open-weight models spend far less than the frontier, averaging 7.2\M tokens and 102 tool calls per instance against 20.9\M and 355 (\autoref{tab:rq1-main}), yet solve under 4\% of instances.
Within a model, failed runs cost more than successes, but across models lower effort tracks weaker capability, and two signals show that lower effort accompanies earlier failure to reach dynamic evidence rather than greater cost efficiency.
First, open-weight runs exhaust the 90-minute budget on only 4.6\% of instances against 23.3\% for the closed-weight configurations, so they stop while time remains rather than grinding to the limit.
Second, only 9 of the open-weight configurations' 2{,}179 illegal PoCs produce an off-target or wrong-class crash on the vulnerable image, compared with 85 of 1{,}298 for the closed-weight configurations (\autoref{fig:failure-modes}).
This lower effort coincides with failure to reach a crash signal, not more efficient target resolution.

\PP{Failures rarely reach any crash signal}
\autoref{fig:failure-modes} partitions the 3{,}486 \cc{illegal} verdicts into the dominant failure modes, of which only the crashing ones would fool the vulnerable-only grader \Jv.
Most illegal PoCs never produce a crash signal rather than crashing at the wrong place, and on the browser engines the \emph{No crash} and \emph{Generic JS} modes together account for over 95\% of all illegal verdicts.
A \emph{Generic JS} failure aborts on an ordinary JavaScript exception, typically an unfinished draft that throws before reaching engine code, and the analogous \emph{No KASAN} mode dominates on Linux.

\PP{Precision-first versus breadth-first search}
\codex GPT-5.5 uses a precision-first policy that probes locally and submits once its evidence supports attribution, solving 48 of its 49 verified V8 instances with a single PoC.
Its illegal pool is small but carries the benchmark's highest off-target share (37.3\% on SpiderMonkey against 3.6\% for \claude in \autoref{fig:failure-modes}), so its failures still reach near-target crashes, at the cost that an instance whose trigger eludes the model yields no submission at all.
In contrast, \claude follows a breadth-first policy, submitting 927 candidate PoCs on V8 against 100 for \codex as it revises a PoC into a chain of numbered scripts, and although 99.1\% of its V8 illegal verdicts never crash the engine, this breadth still reaches target-aligned crashes a single confident attempt misses, solving 9 SpiderMonkey instances that \codex GPT-5.5 misses.

\subsection{RQ4: Real-World Zero-Day Discovery}
\label{s:zero-day}

\begin{table}[t]
  \centering
  \caption{Vulnerabilities surfaced during \sys.
  \emph{Changes} reports added and removed lines in the upstream fix.}
  \label{tab:zerodays}
  \begin{tabularx}{\columnwidth}{@{}l >{\raggedright\arraybackslash}X l c@{}}
  \toprule
  \textbf{ID} & \textbf{Type} & \textbf{Status} & \textbf{Changes} \\
  \midrule
  \RC{01} &
  \makecell[l]{[V8] Sandbox Bypass (exploitable)\\BigInt heap overflow} &
  \makecell[l]{\cmark~Fixed\\\$20,000 bounty} &
  \textcolor{darkgreen}{+77}/\textcolor{red}{-22} \\
  \RC{02} &
  \makecell[l]{[V8] TOCTOU in\\IterableForEach JS execution} &
  \makecell[l]{\cmark~Fixed\\No bounty} &
  \textcolor{darkgreen}{+778}/\textcolor{red}{-254} \\
  \RC{03} &
  \makecell[l]{[SpiderMonkey] Type Confusion in\\\cc{js::AsyncFromSync}\cc{IteratorMethod}} &
  \makecell[l]{\cmark~Fixed\\Duplicate} &
  \textcolor{darkgreen}{+16}/\textcolor{red}{-2} \\
  \bottomrule
\end{tabularx}

\end{table}

Our task names only the target vulnerability type and source files to review, so agents synthesize their own inputs and some yield valid but unintended PoCs that violate memory safety on the latest image.
Three-image grading flags these cases, since a crash on the fully patched build may not be the known, already-fixed vulnerability, so the same workflow that measures agent capability also surfaces zero-day vulnerabilities.
Our runs surface two zero-day vulnerabilities in V8 and a duplicate bug in SpiderMonkey, summarized in \autoref{tab:zerodays}.
We classify \RC{01} as critical because a \cc{BigInt} heap overflow lets a PoC access raw pointers inside the V8 sandbox meant to contain heap corruption~\citep{gross2024v8sandbox}.
We drive the PoC to instruction-pointer control and arbitrary code execution, and the Chrome Vulnerability Reward Program accepted it as an exploitable finding with a \$20{,}000 bounty~\citep{chromium490769268}.
All three targets reward such findings~\citep{chromevrp,mozbugbounty,kernelctf}.

\subsection{Threats to Validity}
\label{s:threats}

The main threat to internal validity lies in the LLM-based judge.
We mitigate it by retaining the three-image evidence for every judged candidate and recording all manual resolutions and definitive-verdict audit corrections before aggregating the final outcomes.
The main threat to external validity is that the covered vulnerabilities may not represent the full spectrum of real-world ones, which we mitigate by collecting 344 validated vulnerabilities from widely deployed browsers and the Linux system, covering a variety of critical vulnerability families.

\section{Discussion}
\label{s:disc}

\PP{Online information leakage}
The default \codex configuration reaches the public internet through both a built-in web-search action and ordinary shell commands, and this access lets the agent recover the exact artifacts our task design withholds.
For example, our earlier network-enabled, root-graded runs solve 122 of 137 Linux instances with \codex GPT-5.5 and 105 with GPT-5.4.
Please note that the final harness enforces per-instance privileges and revised grading, so these scores do not form a matched network ablation.

The trajectories document retrieval of withheld, target-specific evidence.
CVE-2025-38001 opens the exact upstream fix commit, CVE-2025-38177 downloads its exact patch, and GPT-5.4 on CVE-2022-50367 downloads the original syzkaller reproducer.
These artifacts reveal the root cause or trigger and bypass the source-based reconstruction that withholding the reference PoC is designed to measure.
We therefore report Linux \codex results from fresh, fully offline, privilege-aware runs with provider-side web search disabled and no completed shell network commands in their trajectories.

\PP{Privilege-aware validation}
A kernel PoC that reproduces a bug only from a root shell measures something weaker than the CVE claims when the underlying flaw is reachable by an unprivileged user, so grading every PoC at root would conflate the two.
We therefore run each Linux PoC at the privilege the instance's threat model assumes and credit a user-reachable bug only when it crashes without root.
Replaying the PoCs credited by a root-only evaluation at uid 1000 shows that 49 of GPT-5.5's 73 successes on user-labeled instances and 43 of GPT-5.4's 46 do not reproduce without root.

The final privilege-aware runs still solve 78 of the 98 user-labeled instances with GPT-5.5 and 48 with GPT-5.4.
Source inspection shows that 68 of GPT-5.5's 78 successful PoCs on these instances and 43 of GPT-5.4's 48 invoke \cc{unshare(CLONE\_NEWUSER)}.
These PoCs explicitly create a user namespace during uid 1000 execution instead of relying on a root shell.
The remaining 39 instances carry a root label because their reference trigger requires an init-namespace capability, such as mounting a crafted image or loading a module.
We grade these instances at root and apply uid 1000 execution to the 98 user-labeled instances.
The grader therefore rejects a root shortcut only when the instance requires user-level reachability.

\section{Conclusion}
\label{s:conclusion}

We present \sys, a benchmark for agentic bug hunting built from 344 reproducible instances across V8, SpiderMonkey, and the Linux kernel, together with three-image grading that recovers valid PoCs discarded by simpler graders.
Our extensive study of six frontier models and three coding agents reveals findings on their capability, cost, and long-context behavior that we expect to benefit future research on security agent evaluation and training.
Our runs also surfaced three vulnerabilities and earned a \$20,000 Google VRP bounty.

\ificse
\def\BibTeX{{\rm B\kern-.05em{\sc i\kern-.025em b}\kern-.08em
T\kern-.1667em\lower.7ex\hbox{E}\kern-.125emX}}
\IEEEoverridecommandlockouts
\bibliographystyle{IEEEtranS}
{\footnotesize
\bibliography{p,conf}}
\else
\bibliographystyle{sty/colm/template}
\bibliography{p,conf}
\fi

\end{document}